\documentclass[english]{aa}
\usepackage{graphicx}
\usepackage{txfonts}
\usepackage{subcaption}
\begin{document}

\title{N-body simulations of planet formation via pebble accretion I: First Results}

\author{ Soko Matsumura\inst{1} \and Ramon Brasser\inst{2} \and Shigeru Ida\inst{2} }

\institute{School of Engineering, Physics, and Mathematics, University of Dundee, DD1 4HN, UK\\
           \email{s.matsumura@dundee.ac.uk}
           \and
           Earth-Life Science Institute, Tokyo Institute of Technology, Meguro-ku, Tokyo, 152-8550, Japan}

%


\abstract
{
Planet formation with pebbles has been proposed to solve a couple of long-standing issues 
in the classical formation model.
Some sophisticated simulations have been done to confirm the efficiency of pebble accretion. 
However, there has not been any global N-body simulations that compare the outcomes of 
planet formation via pebble accretion with observed extrasolar planetary systems.
}
{
In this paper, we study the effects of a range of initial parameters of planet formation via pebble accretion, 
and present the first results of our simulations.
}
{
We incorporate the pebble accretion model by \cite{Ida16a} in the N-body code SyMBA \citep{Duncan98}, 
along with the effects of gas accretion, eccentricity and inclination damping and planet migration in the disc.
}
{
We confirm that pebble accretion leads to a variety of planetary systems, but 
have difficulty in reproducing observed properties of exoplanetary systems, such as 
planetary mass, semimajor axis, and eccentricity distributions. 
The main reason behind this is a too-efficient type I migration, which sensitively depends on the disc model. 
However, our simulations also lead to a few interesting predictions. 
First, we find that formation efficiencies of planets depend on the stellar metallicities, not only for giant planets, 
but also for Earths (Es) and Super-Earths (SEs).
The dependency for Es/SEs is subtle. Although higher metallicity environments lead to faster formation 
of a larger number of Es/SEs, they also tend to be lost later via dynamical instability.
Second, our results indicate that a wide range of bulk densities observed for Es and SEs is a natural consequence of 
dynamical evolution of planetary systems. 
Third, the ejection trend of our simulations suggest that one free-floating E/SE may be expected for 
two smaller-mass planets. 
}
{} 

\maketitle

%
\section{Introduction}\label{intro}
Recent studies on planet formation have highlighted the pontential importance of 
the effects of cm- to m-sized objects called pebbles.
In particular, pebbles could resolve two main problems in the classical planet formation scenario --- 
(1) formation of planetesimals, and (2) long formation time scales of protoplanetary cores to start gas accretion.  

The first problem is related to the difficulty of forming km-sized objects via collisional agglomeration.
The initial stage of the core accretion scenario proceeds as dust grains collide with one another and grow.  
However, this pairwise growth is expected to stall once the pebble-size ($1~{\rm mm} - 1~{\rm cm}$) 
is reached, due to bouncing and fragmentation \citep[e.g.,][]{Blum08,Brauer08,Zsom10,Windmark12,Birnstiel12a}.  
Even if the objects managed to grow beyond this size, there is a well-known problem called 
a metre-size barrier \citep{Adachi76,Weidenschilling77}, which occurs due to the difference in orbital speed of a gas disc and 
dust particles.
Since the protoplanetary discs rotate at sub-Keplerian velocities due to gas pressure support, 
the dust particles feel a headwind as they orbit faster than the disc; this removes their angular momentum 
and leads to the inward migration.  
This radial speed is the fastest for roughly m-sized objects and 
the migration time scale is about 100 yrs at 1 AU \citep{Adachi76,Weidenschilling77} for typical disc parameters.
 
These rapidly migrating pebbles could lead to planetesimal formation by bypassing the metre-size barrier, 
either via the streaming instability \citep[SI,][]{Youdin05,Johansen07a,Johansen11}, 
or via the gravitational instability \citep[GI,][]{Goldreich73,Youdin02}.
In the SI model, the migrating pebbles could be trapped at local pressure bumps, which are created 
by the magneto-rotational instability (MRI) turbulence \citep[e.g.,][]{Balbus91}
\footnote{Recent studies of protostellar discs suggest that the MRI turbulence may not be efficient 
in planet-forming region ($\sim1-10\,$AU), and that the angular momentum transfer may be largely 
done by magnetocentrifugal disc winds \citep{Turner14}. 
If this were the case, the pressure bumps would need to be created by other mechanisms, 
or pebbles would need to be trapped by other means such as vortices \citep[e.g.,][]{Barge95,Raettig15}.}, 
and gravitationally collapse to form $\sim100-1000\,$km-sized planetesimals \citep{Johansen07b,Johansen11}.
It has been shown that the SI may be difficult to occur because the mechanism 
requires a large solid-to-gas ratio or an extreme disc condition \citep{Krijt16,Ida16b}.
However, the SI could still be possible for some favourable conditions in terms of fragmentation threshold 
velocity and disk properties \citep{Laibe14,Drazkowska16} or just beyond the snow line \citep{Schoonenberg17ap}.  

In the GI model, the planetesimals could form directly as the particles pile up in the inner disc \citep{Youdin02}. 
However, the mechanism is difficult to work except for just inside the snow line \citep{Ida16b}.
Yet another potential channel of planetesimal formation may be direct collisions. 
\cite{Okuzumi12}, for example, showed that formation of planetesimals via collisions of sub-micron-sized porous pebbles 
is possible beyond the snow line, if collisional fragmentation can be ignored. 

Once the planetesimals are formed, they would be exposed to the flux of pebbles, 
which results in a potential solution to the second problem of planet formation.
The suceptibility of pebbles to gas drag, which led to the metre-size barrier for planetesimal formation, 
now helps the growth of protoplanetary cores. 
When planetary embryos are large enough to be in the settling regime $\gtrsim100-1000\,$km \citep{Ormel10,Guillot14}, 
their collisional cross sections become significant.
\cite{Kretke14} showed that the cross-section for the accretion of pebbles is a few orders of magnitude larger than that 
for km-sized planetesimals, and even larger than the Hill radius of the embryo.
This leads to a significant speed-up of planetary growth with pebble accretion 
compared to the classical oligarchic growth \citep{Kokubo00}. 
\cite{Ida16a} estimated that the accretion time scale for an embryo 
due to the three-dimensional mode of pebble accretion has no dependence on mass ($t_{\rm acc}\propto M_{pl}^0$), 
and thus is much faster than the oligarchic growth \citep[$t_{\rm acc}\propto M_{pl}^{1/3}$, see][]{Kokubo00}, 
while the two-dimensional mode has the dependence comparable to the oligarchic growth.  
Therefore, the pebble accretion is likely to replace the oligarchic growth stage until the pebble flux is exhausted. 

The first global simulations of pebble accretion were performed by \cite{Kretke14}, 
who discovered that the pebble accretion was too efficient and $\sim100$ Earth/Mars-like planets could be formed 
if pebbles existed from the beginning.
The follow-up work by \cite{Levison15a} highlighted the importance of evolution of the pebble flux. 
They showed that, by introducing pebbles over some period of time, planetesimals had enough time to 
scatter one another and a system produced a more reasonable number of planets.  

When the pebbles are small, they grow in-situ.  
In this so-called ``drift-limited'' growth regime, pebbles start migrating once the growth time scale becomes comparable 
to the migration time scale \citep{Birnstiel12a,Ida16a}.   
Since the growth time scale increases with radius, the pebbles are formed in a relatively narrow region of a disc, 
and this pebble front moves outward with time \citep{Lambrechts14b,Sato16,Ida16a}. 
Once the pebble front reaches the outer edge of the disc and the solids are exhausted, the pebble flux 
decreases quickly \citep{Sato16,Chambers16}, which is the end of pebble accretion.  
In \cite{Sato16}, this time scale is about several $10^5\,$yr for the disc's outer radius of 100\,AU and about a few Myr for 300\,AU.
Furthermore, the pebble flux into the inner region would be halted once a planet acquires the pebble isolation mass 
\citep{Morbidelli12,Lambrechts14a}. 
In the Solar System, Jupiter and Saturn are considered to have reached these limits but not ice giants \citep{Lambrechts14a}, 
which may also explain why terrestrial planets are poor in water \citep{Morbidelli16}.

The numerical simulations of pebble accretion have been performed by a few different groups. 
\cite{Kretke14} implemented the effects of gas drag into LIPAD, a particle-based Lagrangian code that can follow 
collisional/accretional/dynamical evolution of a protoplanetary system \citep{Levison12}. 
They demonstrated that both giant planets and terrestrial planets, as in the Solar System, can be formed 
via pebble accretion \citep{Levison15a,Levison15b}. 
\cite{Bitsch15} considered formation of a single planet via pebble accretion in a sophisticated disc model, 
and showed where various kinds of giant planets could be formed.
%
\cite{Chambers16} developed a model where pebbles are created via collisions of $\mu$m-sized dust grains 
as well as collisional fragmentation.
He showed that multiple gas giants could be formed within $3\,$Myr, as opposed to classical planet formation simulations 
where embryos barely became large enough to accrete gas within $5\,$Myr. 
He also showed that there were two general outcomes of planet formation depending on the efficiency of pebble accretion. 
When the pebble accretion is efficient, multiple gas giants form beyond the snow line and smaller planets form closer to the star. 
When the pebble accretion is inefficient, on the other hand, no giant planets form and planets remain comparable to or smaller than Earth. 

Most previous simulations, however, assumed a relatively simple disc structure in estimating pebble accretion rates.  
As shown in \cite{Ida16a}, the architecture of planetary systems formed by pebble accretion is expected to be sensitively 
dependent on the disc parameters, for example, radiation dominated or viscous-heating dominated regimes. 
In this paper, we numerically study planet formation in protostellar discs via pebble accretion by 
changing parameters such as the stellar metallicity, the disc mass, and the disc's viscosity.  
We have employed a symplectic N-body integrator SyMBA \citep{Duncan98}, which was modified to include pebble accretion, 
gas accretion, eccentricity and inclination damping, type I and type II migration, and the effects of sublimation.
Instead of following a large number of particles as in \cite{Levison15a}, 
we adopt the analytical model by \cite{Ida16a} to calculate a pebble mass accretion rate onto an embryo. 
This allows us to perform simulations by using wide ranges of parameters and compare the resulting planetary distributions 
with observations.
In Section~2, we introduce our code and initial conditions.  
In Section~3, we present the results without and with planet migration effects.
In Section~4, we discuss and summarise our results.
%
%
%
%
\section{Numerical methods and initial conditions}\label{methods}
To study planet formation with pebble accretion, we perform a large set of numerical simulations 
of a Sun-like star and planetary embryos. 
The planetary embryos are assumed to accrete pebbles and gas, and undergo 
migration and eccentricity and inclination damping from the gas disc.   
A planetesimal disc can be included but is not done so for this work. 

The integrations use the Kepler-adapted symplectic $N$-body code SyMBA \citep{Duncan98}, 
a descendant of the original techniques of \cite{Wisdom91} and \cite{Kinoshita91}. 
As described below, the code has been modified to include the effects of pebble accretion (Section~\ref{pebmodel}), 
gas envelope accretion (Section~\ref{gasaccmodel}) as well as   
eccentricity and inclination damping from the gas disc and planet migration through torques induced 
by the disc-planet interaction (Section~\ref{migmodel}).   
The disc's model we adopted is introduced in Section~\ref{discmodel}, and the initial conditions are described in Section~\ref{ICs}.

The integration step in SyMBA has been modified to include these mechanisms as
\begin{equation}
{\mathcal P}^{\, \tau/2} {\mathcal M}^{\,\tau/2} \, {\mathcal I}^{\,\tau/2} \, {\mathcal D}^{\,\tau} \, {\mathcal I}^{\,\tau/2} \, 
{\mathcal M}^{\,\tau/2} {\mathcal P}^{\, \tau/2} \ , 
\label{eq:step}
\end{equation}
where $\tau$ is the time step and each term is an operator. 
The operator ${\mathcal D}$ advances the planets along their osculating 
Kepler orbits, while ${\mathcal I}$ handles the secular interactions between the planets \citep{Duncan98}. 
Both of these operators function in the democratic heliocentric coordinates (the velocities are barycentric), 
as described in \cite{Duncan98}.
The other two operators function in heliocentric coordinates before and after coordinate transformation. 
Specifically, ${\mathcal M}$ generates radial migration and eccentricity and inclination damping, and ${\mathcal P}$ is 
associated with the accretion of pebbles and gas. 
Even though we do not include planetesimals in our simulations, the code is written such that any planetesimals would not feel the ${\mathcal P}$ operators unless they exceed a specific size, and would feel 
gas drag rather than migration torques for the ${\mathcal M}$ operator.

For these $N$-body simulations, two additional minor adjustments to the SyMBA algorithm were implemented 
following the Swift package\footnote{Available at http://www.boulder.swri.edu/\~{}hal/swift.html}. 
First, SyMBA decomposes the Hamiltonian in canonical heliocentric coordinates 
(i.e., heliocentric positions and barycentric velocities) and employs a multiple time step technique 
to handle close encounters. 
The time step subdivision to handle close encounters is achieved in the SyMBA algorithm by using a 
partition function to decompose the $r^{-2}$ gravitational force between two planets into forces that 
are non-zero only between two cut-off radii, $R_{k+1} \le r < R_{k}$. 
The simplest partition function is the ($2\ell+1$)-th order polynomial in $x$ that has 
$f_\ell(0) = 1$, $f_\ell(1) = 0$, and all derivatives up to the $\ell$-th derivatives zero at $x = 0$ and $1$. 
\cite{Duncan98} found that the third-order polynomial $f_1(x) = 2 x^3 - 3 x^2 + 1$ worked well for many situations, 
which did not include repeated encounters on orbital time scales over $100\,$Myr--$1\,$Gyr. 
For this more challenging situation a smoother partition function is needed. 
We followed \cite{Brasser15} and implemented the next appropriate polynomial 
$f_3(x) = 20 x^7 - 70 x^6 + 84 x^5 - 35 x^4+ 1$. 
We found that the use of $f_3(x)$ sufficed for our needs.

Second, the accretion of pebbles and gas modifies the Hill spheres and physical radii of the planets, 
so these need to be updated at regular intervals. 
Doing so makes the code no longer symplectic, but since the mass of the planets changes slowly enough with time, 
the changes are adiabatic and the system is approximately symplectic. 
We computed the planetary radii using the description of 
\cite{Seager07} for masses below 5~$M_E$, and is given by

\begin{equation}
\log\left(\frac{R_{pl}}{3.3\,R_E}\right)=-0.209+\frac{1}{3}
\log\left(\frac{M_{pl}}{5.5\,M_E}\right)-0.08\left(\frac{M_{pl}}{5.5\,M_E}\right)^{0.4},
\label{eq_MR}
\end{equation}
where $R_{pl}$ and $M_{pl}$ are the radius and mass of the planetary embryo. 
This relation fits Mars, Venus and Earth well. 
For masses in excess of 5~$M_E$, we used the $R_{pl}/R_E = 1.65(M_{pl}/5 M_E)^{1/2}$, 
which fits Jupiter and is acceptable for Uranus and Neptune.

Simulations are run first for 4.6~Myr with a time step of $3\times10^{-4}\,$yr. 
Bodies are removed when they are closer than 0.03~AU or farther than 100~AU from the central star, 
and when they collide. 
We assume perfect accretion during collisions and thus fragmentation effects are not included. 
When $\dot{M}_* < 10^{-9} \, {\rm M_{\odot}/yr}$, we remove the disc away in 500~kyr to mimic 
the photoevaporation effect \citep[e.g.,][]{Alexander14}. 
Afterwards, we simulate the resulting systems upto 50~Myr with SyMBA, by using the same time step and removal criteria.

\subsection{Disc Model}\label{discmodel}
For this study, we adopt the same disc model as \cite{Ida16a}.
We assume a steady accretion rate of the disc gas onto the central star, and thus the accretion rate is
written as: 
\begin{equation}
\dot{M}_* = 3\pi\Sigma_g\nu = 3\pi\alpha\Sigma_gh_g^2\Omega \ ,
\label{eq_msdot}
\end{equation}
where $\Sigma_g$ is the gas surface mass density, $h_g$ is the pressure scale height of a gas disc, and 
$\Omega$ is the orbital frequency. 
The viscosity parameter $\alpha$ is assumed to be constant throughout the disc, 
where the viscosity is written as $\nu=\alpha c_sh_g$ \citep{Shakura73}. 
The disc scale height is related to the temperature and the sound speed via $c_s=h_g\Omega$. 
In this paper, we use the isothermal sound speed instead of the adiabatic one, 
and adopt the standard definition of $c_s^2=k_BT/(\mu m_H)$, where $k_B$ is the Boltzmann constant, 
$\mu=2.34$ is the mean molecular weight, and $m_H\sim1.67\times10^{-24}\,{\rm g}$ is the hydrogen mass.

As shown below and in Section~\ref{pebmodel}, both the disc evolution and the pebble mass accretion rate 
onto an embryo are regulated by the stellar mass accretion rate. 
Throughout the simulations, we change the stellar mass accretion rate as follows \citep{Hartmann98}:
\begin{equation}
\log\left(\frac{\dot{M}_*}{\rm M_{\odot}\,yr^{-1}}\right) = -8 -\frac{7}{5}\log\left(\frac{t}{\rm Myr} + 0.1\right) \ .
\label{eq_M*}
\end{equation} 
Here, the extra $0.1\,$Myr is added to avoid the logarithmic singularity \citep{Bitsch15}.

The disc temperature is mostly determined by the heating source. 
Generally, viscous heating dominates the inner region of the disc while stellar irradiation dominates the thermal 
structure farther out \citep[e.g.,][]{Chambers09}.
The disc midplane temperature can be approximated by $T=\max(T_{\rm vis},\,T_{\rm irr})$, 
where $T_{\rm vis}$ and $T_{\rm irr}$ 
are temperatures in viscous and irradiation heating regions, respectively. 
The disc models by \cite{Garaud07} and \cite{Oka11} are empirically fitted by 
\begin{eqnarray}
T_{\rm vis} &=& 200\,M_{*0}^{3/10}\alpha_3^{-1/5}\dot{M}_{*8}^{2/5}\left(\frac{r}{\rm AU}\right)^{-9/10} \, {\rm K} \nonumber \\
T_{\rm irr} &=& 150\,L_{*0}^{2/7}M_{*0}^{-1/7}\left(\frac{r}{\rm AU}\right)^{-3/7} \, {\rm K} \ ,
\label{eq_temp}
\end{eqnarray}
where $r$ is the distance to the central star and the exponents are derived by analytical arguments. 

Throughout this paper, we use the following normalised parameters 
for the viscosity parameter $\alpha$, stellar luminosity $L_*$, stellar mass $M_*$, 
stellar mass accretion rate $\dot M_*$, and the gas to dust surface mass density $\Sigma_{dg}=\Sigma_d/\Sigma_g$.
\begin{eqnarray}
&&  \alpha_{3} \equiv \frac{\alpha}{10^{-3}}, \ L_{*0} \equiv \frac{L_*}{L_{\odot}}, \ M_{*0} \equiv \frac{M_*}{M_{\odot}} \nonumber \\
&& {\dot M_{*8}} \equiv \frac{\dot M_*}{10^{-8}\,{\rm M_{\odot} \ yr^{-1}}}, \ \Sigma_{dg2} \equiv \frac{\Sigma_{dg}}{0.01},
\end{eqnarray}

With these temperature profiles, we can compute the reduced scale height $h_g/r$ as follows. 
\begin{eqnarray}
\hat{h}_{\rm g,vis} &=& \frac{h_{\rm g,vis}}{r}
\sim 0.027 M_{*0}^{-7/20}\alpha_3^{-1/10}\dot{M}_{*8}^{1/5}\left(\frac{r}{\rm AU}\right)^{1/20} \nonumber \\ 
\hat{h}_{\rm g,irr} &=& \frac{h_{\rm g,irr}}{r}
\sim 0.024 L_{*0}^{1/7}M_{*0}^{-4/7}\left(\frac{r}{\rm AU}\right)^{2/7} 
\label{eq_hr}
\end{eqnarray} 
Therefore, the disc flares in the irradiative region but not in the viscous region.
In this paper, all the quantities with ``\char`\^'' are normalised by the orbital radius $r$, unless 
it is noted otherwise.

Equations~\ref{eq_msdot}, \ref{eq_temp}, and \ref{eq_hr} can be combined to 
compute the gas surface mass density as follows. 
\begin{eqnarray}
\Sigma_{\rm g,vis} &=& 2.1\times 10^3 M_{*0}^{1/5}\alpha_3^{-4/5}\dot{M}_{*8}^{3/5}
\left(\frac{r}{\rm AU}\right)^{-3/5} \, {\rm g\, cm^{-2}} \nonumber \\
\Sigma_{\rm g,irr} &=& 2.7\times 10^3 L_{*0}^{-2/7}M_{*0}^{9/14}\alpha_3^{-1}\dot{M}_{*8}
\left(\frac{r}{\rm AU}\right)^{-15/14} \, {\rm g\, cm^{-2}} 
\end{eqnarray}
The boundary between the viscous and irradiation regions occurs at 
$r_{\rm vis,irr} \simeq 1.8L_{*0}^{-20/33}M_{*0}^{31/33}\alpha_3^{-14/33}\dot{M}_{*8}^{28/33}\,$AU. 

In the simulations, the disc temperatures and surface-mass densities evolve with time, as the stellar mass accretion rate 
$\dot{M}_{*8}$ decreases.  However, we keep their gradients the same.  
We do not consider opacity variations in this work either and adopt a constant opacity of $\kappa\sim1\,{\rm cm^2\,g^{-1}}$. 

\begin{figure*}[ht]
\begin{subfigure}{.5\textwidth}
\includegraphics[width=1.\linewidth]{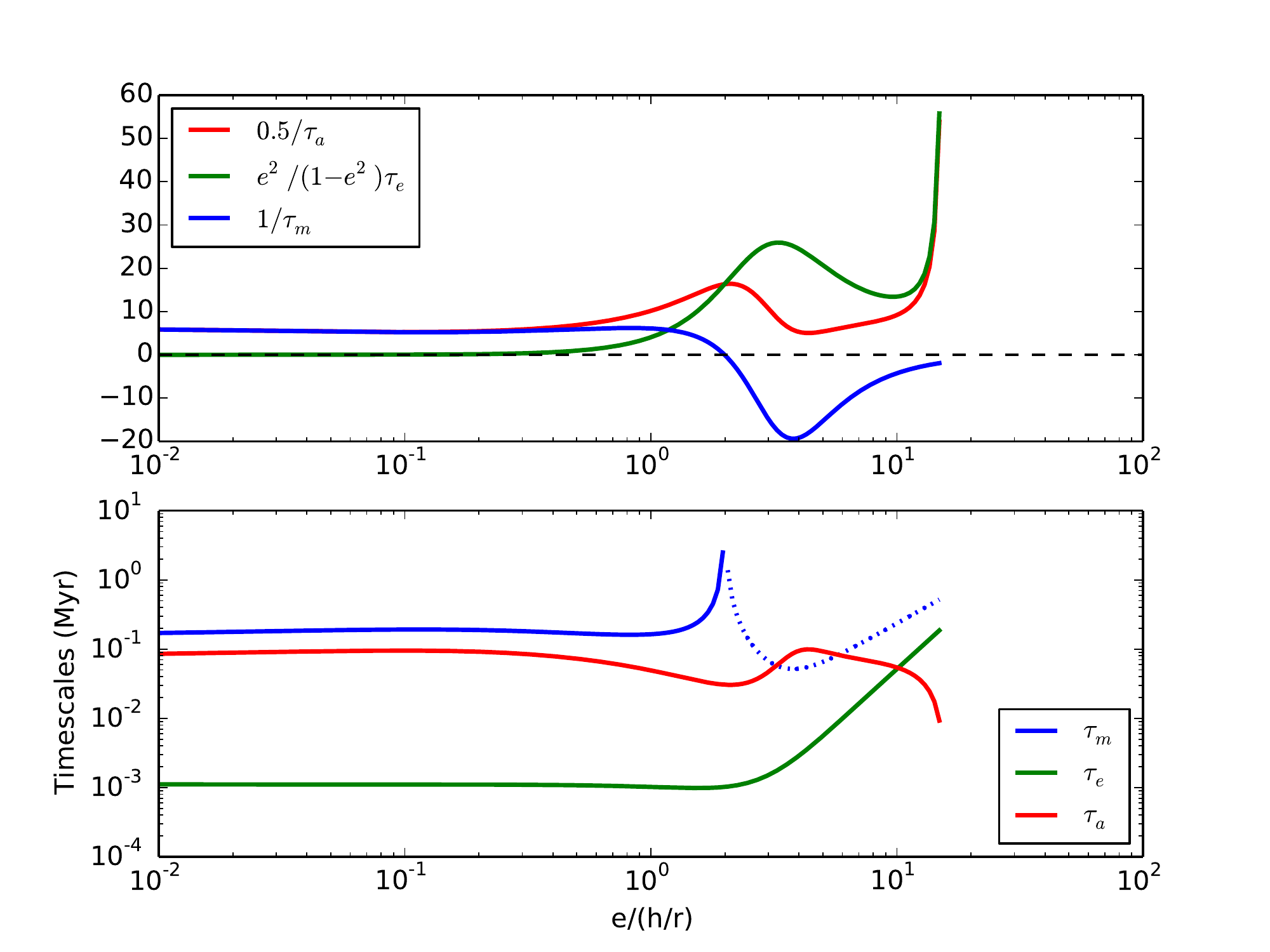}
\end{subfigure}
\begin{subfigure}{.5\textwidth}
\includegraphics[width=1.\linewidth]{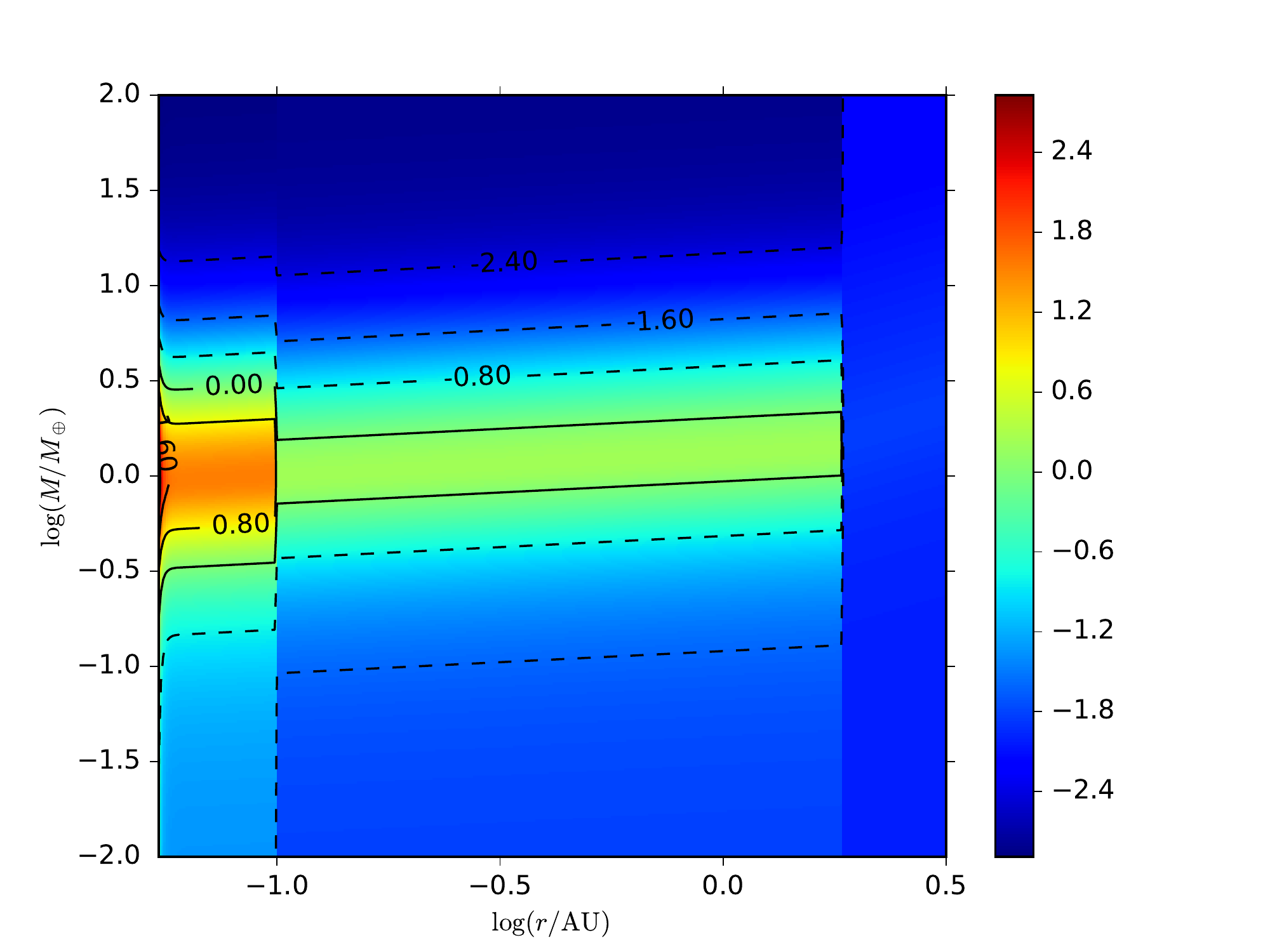}
\end{subfigure}
\caption{Left panel, Top: comparing terms in Eq.~\ref{eq_tauaem} for 1 Earth-mass planet. 
Left panel, Bottom: corresponding time scales. The dotted line indicates when the ``migration'' time scale $\tau_m$ 
becomes negative, which could lead to an artificial outward migration if $\tau_m$ is used in place of $\tau_a$ in 
Eq.~\ref{eq:mig}.  
Right panel: Normalised torques for various planetary masses in the inner region of a disc. 
The torque becomes positive for an Earth-mass planet. The discontinuity around $0.1\,$AU is caused by 
the treatment of the inner disc edge (see Section~\ref{discmodel}). \label{fig_tauaem}}
\end{figure*}
The inner edge of the disc is treated so that the surface density of the disc is smoothly decreased to zero according to
\begin{equation}
 \Sigma_g = \Sigma_g(r_{\rm tr})\tanh\left(\frac{r_{\rm tr}-r_{\rm in}}{h_g}\right)
\end{equation}
where we set the transition radius at $r_{\rm tr}=0.1\,$AU because it is unclear how reliable the employed disc model is 
closer to the Sun where MRI effects play an increasingly important role. 
The inner edge of the disc is set at $r_{\rm in}=0.062\,$AU, i.e., at the 2:1 mean-motion resonance with $r_{\rm tr}$. 
This prescription causes a discontinuity in the surface density slope at $r_{\rm tr}$, 
which is seen in the right panel of Figure~\ref{fig_tauaem} for the torque contours.

\subsection{Pebble Accretion Rate}\label{pebmodel}
We implement the pebble accretion model by following \cite{Ida16a}. 
Here, we briefly summarise their model and refer the readers to their paper for details.
The pebble mass accretion rate onto a planetary embryo depends on whether the accretion can be 
considered as two- or three-dimensional \citep{Ida16a}, where the transition occurs 
when the collisional cross section radius of pebble flows, $b$, becomes large compared to 
a pebble disc's scale height, $h_p$ (i.e., $b \gg h_p$).  
\cite{Ida16a} wrote the pebble mass accretion rate as follows: 
%
\begin{equation}
{\dot M}_p = \min\left(\dot{M}_{2D},\,\dot{M}_{3D}\right) 
=\min\left(\sqrt{\frac{8}{\pi}}\frac{h_p}{b},\,1\right)\,\sqrt{\frac{\pi}{2}}\frac{b^2}{h_p}\Sigma_p\Delta v \ .
\label{eq_Mpdot}
\end{equation}
This is the mass growth rate for our planetary embryos. 
Here, $\Sigma_p$ is the surface mass density of a pebble disc and $\Delta v$ is the relative speed between an embryo 
and a pebble, which can be written as:
\begin{equation}
\Sigma_p\Delta v \sim \frac{\dot{M}_F}{4\pi r\tau_s}\zeta^{-1}\chi\left(1+\frac{3\hat{b}}{2\chi\eta}\right) \ .
\label{eq_sigpdelv}
\end{equation}  
Here, $\dot{M}_F$ is the pebble mass flux, which is estimated from the dust mass swept by 
the pebble formation front per unit time \citep{Lambrechts14b}:
\begin{eqnarray}
{\dot M}_F &\sim& 2\pi r_{\rm pf} \Sigma_{dg}\Sigma_g(r_{\rm pf})\times \frac{dr_{\rm pf}}{dt} \nonumber \\
           &\simeq& 10^{-4} \Sigma_{dg2}\,\alpha_{3}^{-1}
			L_{*0}^{-2/7}M_{*0}^{9/4}{\dot M_{*8}} \ {\rm M_E \ yr^{-1}} \ .
\label{eq_MFdot}
\end{eqnarray}
The pebble mass flux decreases with time, as the stellar mass accretion rate 
$\dot{M}_{*8}$ decreases and thus the disc's density decreases.   

In Eq.~\ref{eq_sigpdelv}, the term in the bracket is proportional to the relative velocity $\Delta v$, which is expressed 
as the sum of the pebble's drift velocity and a contribution from the Keplerian shear \citep{Ormel10,Guillot14}. 
Here $\eta$ is the difference between gas and Keplerian velocities due to the pressure gradient and given by 
$\eta = \hat{h}_g^2/2|d\ln P/d\ln r| = \hat{h}_g^2/2(q+5/2)$, where $q$ is the power of the disc's scale height \citep[see][]{Ida16a}.

The quantities $\zeta$ and $\chi$ in Eq.~\ref{eq_sigpdelv} are functions of the Stokes number $\tau_s$ (see below) 
and are defined as 
\begin{eqnarray}
\zeta & = & \frac{1}{1+\tau_s^2} \nonumber \\
\chi & = & \frac{\sqrt{1+4\tau_s^2}}{1+\tau_s^2} \ .
\end{eqnarray}

The Stokes number $\tau_s$ in Eq.~\ref{eq_sigpdelv} measures the efficiency of gas drag with respect to 
the orbital time scale:
\begin{equation}
\tau_s=t_{\rm stop}\Omega = \frac{\rho_sR}{\rho_gh_g}\max\left(1,\frac{R}{9/4\lambda_{\rm mfp}}\right) 
\label{eq_taus}
\end{equation}
%
where $t_{\rm stop}$ is a stopping time of a pebble due to gas drag, $\Omega$ is the orbital frequency, 
$\rho_s$ and $R$ are a bulk density and a physical radius of a pebble, 
and $\rho_g \sim \Sigma_g/(\sqrt{2\pi}h_g)$ is a gas disc's density.  
Except for the innermost region of the disc, $\tau_s\ll 1$ and thus the gas drag is efficient 
and the collisional cross section for pebble accretion is large. 
This is the reason why pebble accretion is very efficient \citep{Ormel10,Lambrechts12,Kretke14}. 
We can see from Eq.~\ref{eq_sigpdelv} that the pebble mass accretion rate increases for smaller $\tau_s$.

In Eq.~\ref{eq_taus}, two terms in the bracket correspond to Epstein ($R<9/4\lambda_{\rm mfp}$) 
and Stokes ($R>9/4\lambda_{\rm mfp}$) regimes, where $\lambda_{\rm mfp}$ is the mean free path that is expressed as 
\begin{equation}
\lambda_{\rm mfp} = \frac{\sqrt{2\pi}\mu m_Hh_g}{\sigma\Sigma_g} \, .
\end{equation}
Following \cite{Ida16a}, we adopt $\sigma=2\times10^{-15}\,{\rm cm^{2}}$ as the collisional cross section for ${\rm H_2}$.
In the Epstein regime, the mean free path is larger than the pebble radius and 
thus pebbles can be treated as particles, while in the Stokes regime, pebbles behave as a fluid.
The transition from the Epstein to the Stokes regime tends to occur toward the inner region of a disc: 
$r_{\rm ES} \simeq 2.9L_{*0}^{-3/13}M_{*0}^{17/26}\alpha_3^{-21/52}\dot{M}_{*8}^{21/52}\,$AU 
\citep{Ida16a}.

In implementing the above prescriptions, we further take the following three effects into account. 
First, as described in the introduction, 
pebble accretion proceeds as the pebble formation front moves outward through the disc, 
and accretion occurs from the outside in. 
Therefore, we compute a reduction factor $1-\dot{M}_p/\dot{M}_F$, 
which sucessively decreases the pebble flux after passing by each planet from the outside sunwards.
It is meant to take into account each planet accreting some of the pebbles 
as these spiral towards the central star. 

Second, pebble accretion is halted on an embryo when it reaches the pebble isolation mass, which is
approximately $M_{p,iso} \simeq (1/6)M_H$, and is roughly $20\,M_E$ at 5~AU \citep{Lambrechts14b}.
Here, $M_H$ is the planetary mass required for its Hill radius to be comparable to 
the disc's scale height: $M_H = 3(h_{g,pl}/r_{pl})^3M_*$. 
We assume that, once a planet reaches this mass, it will stop accreting and no more pebbles are allowed to 
flow past it to other planets residing closer to the star.

Third, since volatiles inside pebbles could vaporise within the snow line \citep[e.g.,][]{Saito11}, 
the size of pebbles and properties of pebble accretion would change once the snow line is crossed. 
To take account of this effect, we make a simplified assumption that 
the pebble mass flux is halved inside the snow line 
and that the pebbles fragment into silicate grains of 1~mm in diameter \citep{Morbidelli15}.
The location of the snow line in our disc model is at 
\begin{equation}
r_{\rm snow}\sim \max(1.2M_{*0}^{1/3}\alpha_3^{-2/9}\dot{M}_{*8}^{4/9},\, 0.75L_{*0}^{2/3}M_{*0}^{-1/3}) \, {\rm AU} \ ,
\end{equation} 
where the two terms in the bracket correspond to the viscous and irradiation regions, respectively \citep{Ida16a}.  
%
%
\subsection{Gas envelope accretion}\label{gasaccmodel}
Gas envelope accretion occurs for planets that are massive enough, and when the accretion onto the planetary core is low enough 
for the gas envelope to cool and begin contracting \citep{Ikoma00}. 
The critical core mass above which the gas accretion occurs is \citep{Ida04a,Ikoma00}
\begin{equation}
M_{pl,{\rm crit}} \simeq 10\left(\frac{\dot{M}_{\rm core}}{10^{-6}\,M_E\,{\rm yr}^{-1}}\right)^{1/4}\,M_E \ .
\end{equation}
Here, the core accretion rate in our case corresponds to the pebble mass accretion rate $\dot{M}_{\rm core} = \dot{M}_p$ 
(see Eq.~\ref{eq_Mpdot}). 
We also assume that the grain opacity is $\kappa \sim 1 \, {\rm cm^2\, g^{-1}}$ and 
thus there is no dependence of the accretion rate on the opacity. 
The gas envelope then collapses and accretes onto the core on the Kelvin-Helmholtz time scale \citep{Ikoma00} as
\begin{equation}
\tau_{\rm KH} \simeq 10^{9}\left(\frac{M_{pl}}{M_E}\right)^{-3}\,{\rm yr} \ .
\label{eq_tauKH}
\end{equation}

This runaway gas accretion is limited by how quickly the disc can supply the gas to the planet 
(i.e., the gas accretion rate throughout the disc $\dot{M}_*$).
Furthermore, gas envelope accretion stops roughly when the Hill radius is twice the disc scale height, 
i.e., when $M_{pl}=8M_H$ \citep{DobbsDixon07}.
Taking account of these effects, we write the accretion rate onto the planetary core as follows. 
\begin{equation}
 \dot{M}_g = \min\left[\frac{M_{pl}}{\tau_{\rm KH}},\, \dot{M}_*\exp\left(-\frac{M_{pl}}{8M_H}\right)\right] 
 \label{eq_Mg1}
\end{equation}
We limit this gas accretion to the Bondi accretion rate, which is given by
\begin{equation}
 \dot{M}_{g,B} = \frac{4\pi\rho_gG^2M_{pl}^2}{c_s^3}  \ .
 \label{eq_Mg2}
\end{equation}
We note that this threshold was never reached in our simulations, 
which is consistent with the discussion in \cite{Ida04a}. 
In the numerical simulations, we keep track of how much mass in solids and in gas the planet accretes 
so that core and envelope masses can be estimated separately. 
\subsection{Planet migration}\label{migmodel}
The presence of the gas disc causes the embedded embryos and planets to experience torques and tidal forces that result in a combined 
effect of radial migration and the damping of the eccentricities and inclinations. For low-mass planets that are unable to carve a gap 
in the disc, the migration is of type I \citep{Tanaka02}, 
while massive planets that clear the gap in a disc experience type II migration \citep{Lin86}. 

We follow \cite{Coleman14} and implement the eccentricity and inclination damping as well as planet 
migration as follows.
\begin{eqnarray}
\vec{a}_e &=& -\frac{v_r}{\tau_{e}}\,\hat{\bf r} -\frac{0.5(v_{\theta}-v_K)}{\tau_e}\,\hat{\boldmath \theta} \nonumber \\
\vec{a}_i &=& \frac{2A_{cz}v_z + A_{sz}z\,\Omega}{\tau_{i}}\, \hat{\bf k}, \nonumber \\
\vec{a}_a &=& -\frac{\vec{v}}{\tau_{a}}  
\label{eq:mig}
\end{eqnarray}
%
Here, $\hat{\bf r}$, $\hat{\bf \theta}$, and $\hat{\bf k}$ are unit vectors 
for radial, angular, and vertical directions while $\vec{v}$ is the velocity vector of a planetary embryo.  
Following \cite{Coleman14}, we use $A_{cz}=-1.088$ and $A_{sz}=-0.871$.
We use these equations both in the type I and type II regimes.

The eccentricity and inclination damping time scales are taken from \citep{Cresswell07}, 
where they fitted hydrodynamical simulations as
\begin{eqnarray}
\tau_{e} &=& 1.282\,t_{\rm wav}(1-0.14\hat{e}^2+0.06\hat{e}^3+0.18\hat{e}^2\hat{i}^2) \nonumber \\
\tau_{i} &=& 1.838\,t_{\rm wav}(1-0.30\hat{i}^2+0.24\hat{i}^3+0.14\hat{e}^2\hat{i}^2) \ ,
\label{eq_tauei}
\end{eqnarray}
where $\hat{e} = e/(h_g/r)$ and $\hat{i} = \sin(i)/(h_g/r)$ and 
the characteristic time of the orbital evolution is defined as \citep{Tanaka04}:   
\begin{equation}
t_{\rm wav} = \left(\frac{M_*}{M_{pl}}\right)\left(\frac{M_*}{\Sigma_g r_{pl}^2}\right)
\left(\frac{h_{g,pl}}{r_{pl}}\right)^4\Omega_{pl}^{-1} \ .
\end{equation}
%
Thus, the damping is exponential ($\tau_{e}\propto \hat{e}^0$) for a small eccentricity ($\hat{e}<1$) 
and slower ($\tau_{e}\propto \hat{e}^3$) for a high eccentricity.
A similar relation holds for the inclination. 
We apply the eccentricity and inclination damping only when $\hat{e} > 0.001$ and when $\hat{i}> 0.001$. 

For the last equation of Eq.~\ref{eq:mig}, we use the semimajor axis evolution time scale: $\tau_a = -a/\dot{a}$ 
rather than the ``migration'' time scale that is defined as the angular momentum evolution time scale: 
$\tau_m = -L/\dot{L}$ \citep{Muto11,Tanaka02} to avoid an artificial outward migration at high eccentricities (see below). 
They are related as 
\begin{equation}
\frac{1}{\tau_a} = 2\left(\frac{1}{\tau_m} + \frac{e^2}{1-e^2}\frac{1}{\tau_e}\right) \ .
\label{eq_tauaem}
\end{equation}
In the type I regime, we write the ``migration'' time scale as follows:
\begin{equation}
\tau_{m} = -\frac{t_{\rm wav}}{\Gamma/\Gamma_0}\hat{h}_g^{-2} \ , 
\end{equation}
where $\Gamma/\Gamma_0$ is the normalised torque and negative for inward migration \citep{Paardekooper11}.
With this definition, $\tau_m$ becomes negative for $\Gamma/\Gamma_0>0$ (see Figure~\ref{fig_tauaem}).
For the total torque $\Gamma$, we adopt Eq.~15 of \cite{Coleman14} which 
takes account of the reduction in both Lindblad and corotation torques for planets 
on eccentric or inclined orbits. 

On the left panel of Figure~\ref{fig_tauaem}, the top figure compares terms in Eq.~\ref{eq_tauaem}, while the bottom one 
shows each time scale.  The figure demonstrates that, although the ``migration'' time scale $\tau_m$ becomes 
negative for high eccentricities (which would lead to outward migration 
if used in place of $\tau_a$ in Eq.~\ref{eq:mig}), 
the actual semimajor axis evolution time scale $\tau_a$ stays positive.   
This is because the term including the eccentricity damping time scale $\tau_e$ becomes non-negligible for 
high eccentricities ($e>(h_g/r)$). 
The trend is consistent with hydrodynamical simulations done by \cite{Cresswell07}, where 
they observed no reversal of the migration 
for large eccentricity, but found that the torque became positive for $e\sim0.3$.  
The right panel of Figure~\ref{fig_tauaem} shows the normalised torques for various planetary masses in the 
inner region of a disc.  In the default disc conditions, the torque is expected to become positive 
for an Earth-mass planet.
Once the planet is massive enough to open a gap annulus in the disc in its direct surroundings, type II migration sets in. 
This migration occurs on the viscous evolution time scale of the disc \citep{Lin86,Ward97} 
if the disc interior to the planet is more massive than the planet, 
or on a longer time scale otherwise \citep{Hasegawa13}:  
%
\begin{equation}
\tau_{t2} = \frac{2}{3}\frac{r_{pl}^2}{\nu} \, \max\left(1, \, \frac{M_{pl}}{2\pi\Sigma_gr_{pl}^2}\right) \ .
\label{eq_typeII}
\end{equation}
Here, the former in the bracket is the classical type II migration rate $\tau_{t2}\simeq (2/3)(r_{pl}^2/\nu)$, where 
the planetary mass is smaller than the disc mass interior to its orbital radius (i.e., disc-dominated regime),  
and the latter corresponds to the planet-dominated regime when the planet is pushed by the outer disc 
$\tau_{t2}\simeq M_{pl}/\dot{M}_*$ \citep{Hasegawa13}.

How the transition between type I and type II occurs is not fully understood. 
Here we follow the procedure of \cite{Bate03}, which provided an interpolation formula between the two regimes 
based on hydrodynamical simulations.  
The semimajor axis evolution time scale then becomes
\begin{equation}
 \tau_{\rm a} = \frac{\tau_{\rm t1}}{1+(M_{pl}/M_t)^3}+\frac{\tau_{\rm t2}}{1+(M_t/M_{pl})^3},
 \label{eq:bate1}
\end{equation}
where $M_t=(3/5)M_H$ \citep{Bate03} and $\tau_{\rm t1}$ is $\tau_a$ defined in Eq.~\ref{eq_tauaem}.  
This is ultimately used in the last of Equations~\ref{eq:mig}.  

The prescription for migration from equations (\ref{eq:mig}) may cause artificial inward migration 
for high-mass planets at very low eccentricity, when $\tau_e$ becomes comparable to the orbital period.
To avoid this artificial migration, we use the same interpolation equation as (\ref{eq:bate1}), which becomes
\begin{equation}
 \tau_{\rm e} = \frac{\tau_{\rm e}}{1+(M_{pl}/M_t)^3}+\frac{0.1 \tau_{\rm t2}}{1+(M_t/M_{pl})^3}.
 \label{eq:bate2}
\end{equation}
The factor 0.1 is chosen from numerical experiments 
so that we observe no artificial migration but at the same time 
the damping time scale is shorter than the type II migration time scale.
The prescription is rather arbitrary, but the eccentricity damping for high-mass planets is 
not well understood \citep[e.g.,][]{Kley12}.

\begin{figure*}[ht]
\includegraphics[width=1.\linewidth]{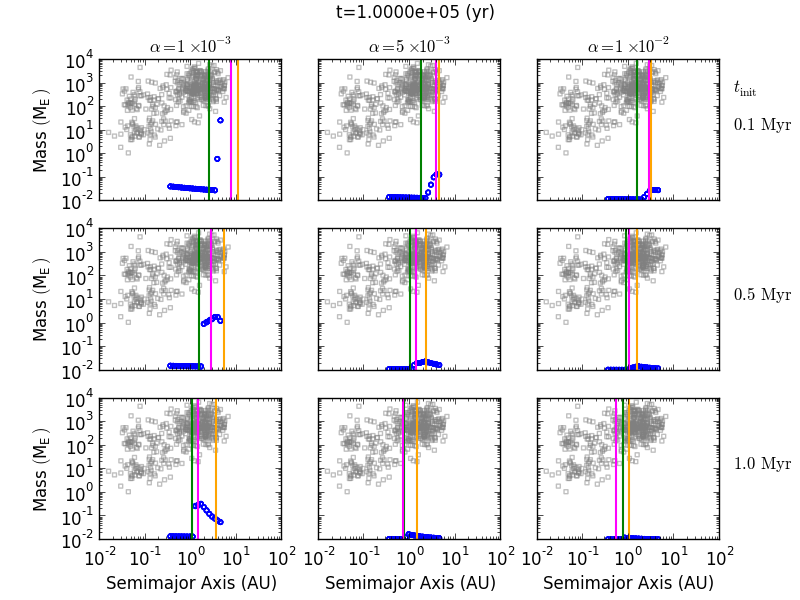}
\caption{The planetary mass-semimajor axis distribution at the simulation time of $0.1\,$Myr 
for a solar metallicity disc with different combinations of a disc's viscosity parameter $\alpha$ and an initial disc age $t_{\rm init}$.  
Here, $t_{\rm init}=0.1$ and $0.5\,$Myr correspond to the inital $t=0$ and $0.4\,$Myr, respectively, in Eq.~\ref{eq_M*}. 
The green, pink, and orange lines indicate the snow line, the boundary between viscous and irradiation regimes, 
and the boundary between Stokes and Epstein regimes, respectively.
Each panel shows the results from five simulations as blue circles, and grey squares represent the corresponding parameters for 
planets observed by the radial velocity method. 
A peak and a jump seen in mass growth can be explained by different regimes of the disc (see text). \label{fig_aMIda16}}
\end{figure*}
\begin{figure*}[ht]
\includegraphics[width=1.\linewidth]{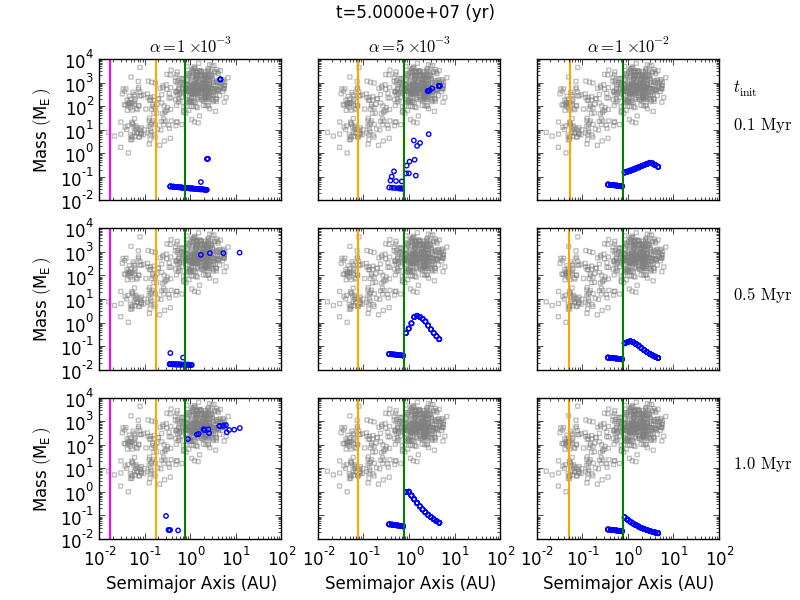}
\caption{The same as Figure~\ref{fig_aMIda16}, but at the simulation time of $50\,$Myr. \label{fig_aMIda16_2}}
\end{figure*}

\subsection{Initial Conditions}\label{ICs}
To explore a variety of planet formation environments, we change the following parameters.  
We vary the stellar metallicity (which changes the dust to gas ratio and thus $\dot{M}_F$) roughly over 
the range of the metallicities of observed planet-hosting stars: ${\rm [Fe/H]}=(-0.5, \, 0.0, \, 0.5)$.  
This is related to the dust-to-gas ratio as follows \citep[e.g.,][]{Ida08a}.
\begin{equation}
\frac{\Sigma_{dg}}{\Sigma_{dg,\odot}} = 10^{\rm [Fe/H]}  \ ,  
\end{equation}
where $\Sigma_{dg,\odot}$ is the dust-to-gas ratio in the protosolar disc, and we assume the value of $0.01$.
We also vary the initial disc age (i.e., the initial disc mass) over $t_{\rm init}=(0.1, \, 0.5, \, 1.0)\,$Myr, which 
corresponds to the initial stellar mass accretion rates of $\dot{M}_{*8}=(25.1, 2.64, 1.0)\,{\rm M_{\odot}/yr}$.
The disc's viscosity parameter is changed over $\alpha=(10^{-3}, \, 5\times10^{-3}, \, 10^{-2})$ 
\footnote{The viscosity parameter could be much smaller, for example, $\alpha=10^{-4}$.  
However, such a low viscosity parameter leads to an unrealistically high initial disc mass in our current disc model.  
Therefore, we focus on the quoted range of parameters for this paper.}.
We run five simulations per combination of above parameters to minimise variations in the outcome.  
Therefore, for one set of simulations, we run 135 cases.
For this study, we run one set of simulations with both type I and type II migrations switched off, and 
another set of simulations with both effects included. 
 
All seed embryos are approximately a lunar mass and spaced 70 Hill radii apart. 
In other words, the semi-major axis of embryo $n$ is 
\begin{equation}
a_n = a_0[1+K(2M_{pl}/(3M_{\odot}))^{1/3}]^n \ ,
\end{equation} 
where $K = 70$ so that the embryo spacing nearly follows a geometric progression. 
This choice is based on the following argument. 
In the absence of migration, as the embryos accrete pebbles, 
the quantity $K(2M_{pl}/(3M_{\odot}))^{1/3}$ stays roughly constant. 
For the system to remain stable for at least 10~Myr requires their mutual spacing to be $\gtrsim7$ 
Hill radii \citep{Chambers96,Kokubo98,Pu15}.
The pebble isolation mass near 5~AU is approximately $10M_E$ so that $K$ has to decrease by an order
of magnitude as the embryos accrete pebbles. Seed embryos are initially distributed between 0.3~AU
and 5~AU, and there are 19 lunar-sized embryos in total with the above-mentioned spacing.
Most N-body simulations of planet formation so far have assumed a uniform radial distribution of planetary embryos, 
because their radial distribution is unclear.  
Our work also follow this treatment and thus the choice of the uniform radial distribution 
of equal-mass embryos is rather arbitrary.  
%
%
%
\section{Results}\label{results}
In this section, we show main results from our simulations both without and with migration.
\subsection{Agreement with \cite{Ida16a}}\label{sec_aMIda16} 
Before we discuss the overall results, we first check the agreement with the work by \cite{Ida16a}.
Figures~\ref{fig_aMIda16} and \ref{fig_aMIda16_2} compare semimajor axis-mass distributions at 0.1 and 50\,Myr 
for no-migration simulations with the solar metallicity (i.e., ${\rm [Fe/H]}=0.0$).
The nine panels represent different combinations of the viscosity parameter $\alpha$ and an initial disc age $t_{\rm init}$, 
and each panel shows the combined outcomes of five simulations.
The upper left panel with the shortest $t_{\rm init}$ and the smallest $\alpha$ 
correspond to the most massive disc, and the initial disc mass decreases 
with both $t_{\rm init}$ and $\alpha$.  
The green, pink, and orange lines indicate the snow line, the boundary between viscous and irradiation regimes, 
and the boundary between Stokes and Epstein regimes, respectively.  
These lines are plotted in the case of 50\,Myr as well for a reference, although the disc has been long gone by that time. 
Grey squares are the corresponding values for the observed planets.  
To minimise the observational biases, we only use the planets detected by the radial-velocity (RV) method in this paper unless 
it is noted otherwise
\footnote{
Transit survey is strongly biased toward close-in planets, microlensing survey is highly biased to $1-3\,$AU, 
and direct imaging detects only widely separated gas giants. Although the RV survey is also biased, the bias 
is much weaker than the other methods.
}.
We use the list of confirmed planets from \url{http://exoplanets.org} that was taken on 11 April 2017.
 
The overall trends seen in the figures reproduce expectations from Figures~2 and 3 of \cite{Ida16a}. 
First, the planetary masses are increased beyond the snow line because we take account of the ice sublimation effect 
of the pebble flow within that.
Next, as shown in their equations~74 and 75, pebble accretion time scales increase with orbital radii for 
most regimes (i.e., the inner planets grow faster than the outer ones), except for 
the viscous and Stokes regime.  
In our figure, this region is the left of both pink and orange lines.
Indeed, in these regions, we observe that planetary masses increase with distance instead of decreasing.    
Since the planetary masses decrease with distance in the other regions, our simulations tend to have the most optimised regions 
for planet formation beyond the snow line, around a few AU. 
This is consistent with the expectation from \cite{Ida16a} (also see discussion in Section~\ref{sec_lowmasspl}).  

Figure~\ref{fig_aMIda16_2} also shows the two types of planetary systems suggested by \cite{Chambers16} --- 
systems with giant planets with low-mass planets interior to them (upper left panels) 
and systems with only low-mass planets ($\lesssim 1\,M_E$, lower right panels). 
However, we should note that our simulations with the highest metallicity (${\rm[Fe/H]}=0.5$, figures not shown here) 
led to formation of systems of Super Earths with all planets having masses $\sim1-10\,M_E$ as well.  
These results further confirm the claim that pebble accretion could lead to a variety of planetary systems 
\citep[e.g.,][]{Levison15b,Bitsch15,Ida16a,Chambers16}.

\subsection{Overall Results}\label{sec_aMall}
\begin{figure*}[ht]
\includegraphics[width=1.\linewidth]{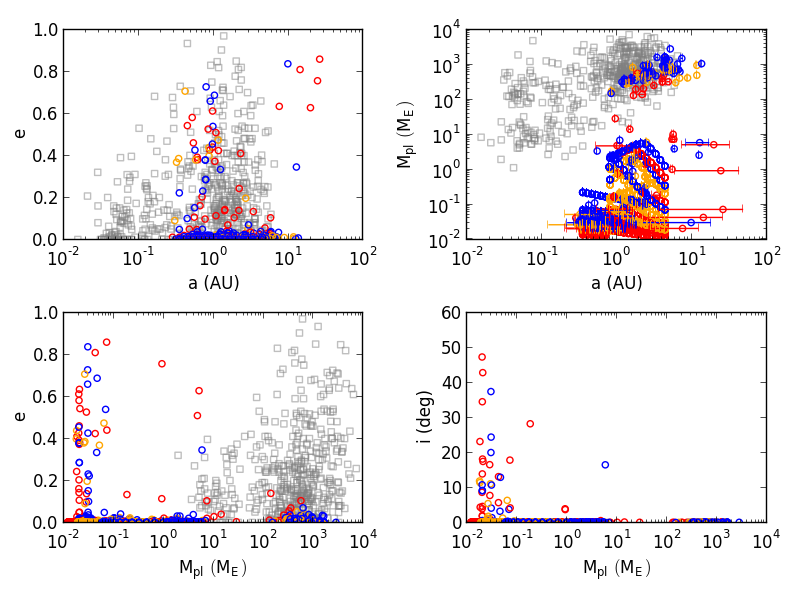}
\caption{Various parameters of planets in no migration simulations at 4\,Myr.  
The top left, top right, bottom left, and bottom right panels show semimajor axis and eccentricity, 
semimajor axis and planetary mass, planetary mass and eccentricity, and planetary mass and inclination, respectively. 
The grey squares in the first three panels are corresponding values for planets observed by the RV method. 
In the top right panel, the error bars indicate pericentres and apocentres.
Also, the blue, orange, and red symbols correspond to stellar metallicities of ${\rm [Fe/H]}=0.5$, 
0.0, and -0.5, respectively.
\label{fig_aeim_nomig1}}
\end{figure*}
\begin{figure*}[ht]
\includegraphics[width=1.\linewidth]{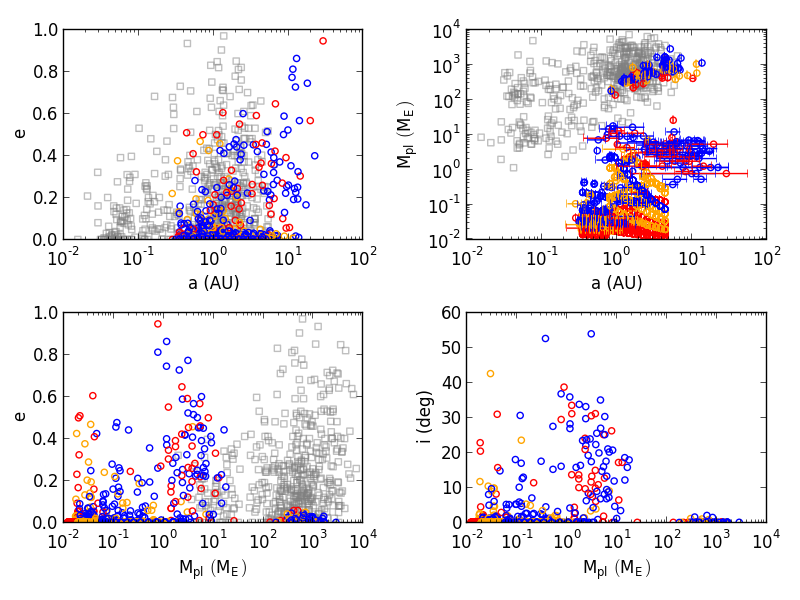}
\caption{The same set of figures as Figure~\ref{fig_aeim_nomig1} for no migration simulations, 
but at the end of the simulations (50\,Myr).
\label{fig_aeim_nomig2}}
\end{figure*}
Figures~\ref{fig_aeim_nomig1} and \ref{fig_aeim_nomig2} show the results of no migration cases just before the 
gas disc dissipation (4\,Myr) and at the end of the simulations (50\,Myr).
Each panel shows the relations between semimajor axis, planetary mass, eccentricity, and inclination 
for all the combinations of $\alpha$ and $t_{\rm init}$.
The blue, orange, and red symbols correspond to stellar metallicities of ${\rm [Fe/H]}=0.5$, 
0.0, and -0.5, respectively. 
The grey squares are corresponding values for planets observed by the RV method. 

The $a-M_{pl}$ distribution of Figure~\ref{fig_aeim_nomig1} (top right) shows that 
all kinds of planets are formed by $4\,$Myr, 
from giant planets ($\geq 0.1\,M_J\sim30\,M_E$) to Earths or Super Earths (Es and SEs from here on, 
$0.1\,M_E-0.1\,M_J$) 
\footnote{
In this paper, we will not distinguish SEs and mini-Neptunes because the envelope mass could vary 
over a few orders of magnitude for the same planetary mass (see Section~\ref{sec_composition}).
}.
In our simulations, the fastest giant planet formation occurs for the massive, low-viscosity, 
and metal-rich discs (see Section~\ref{sec_met}). 

From the $a-e$ distribution (top left panel of Figure~\ref{fig_aeim_nomig1}), 
it appears that some planets achieve quite eccentric orbits despite that they are still embedded in the gas disc. 
However, from the bottom two panels, it is apparent that planets which achieve high eccentricities or inclinations 
are largely limited to small-mass planets ($<0.1\,{\rm M_E}$).
These planets are scattered by larger planets and gain high eccentricities and inclinations, but the damping 
time scale is longer than the disc's lifetime.

Once the gas disc dissipates, dynamical instability kicks in and more massive planets start having 
higher eccentricities and inclinations.
The effects are clearly seen in Figure~\ref{fig_aeim_nomig2}. 
Such an instability leads to a population of eccentric Es/SEs beyond several AU, where we did not have seed embryos.
Giant planets, on the other hand, experience little change in their orbits beyond 
several Myr. 
This is probably because the average number of giant planets we form is $\leq 3$ per system 
for the combinations of parameters that lead to the giant planet formation 
(see Figure~\ref{fig_number} and its discussion in Section~\ref{sec_met}). 
We will come back to this point in Section~\ref{discussion}.

\begin{figure*}[ht]
\includegraphics[width=1.\linewidth]{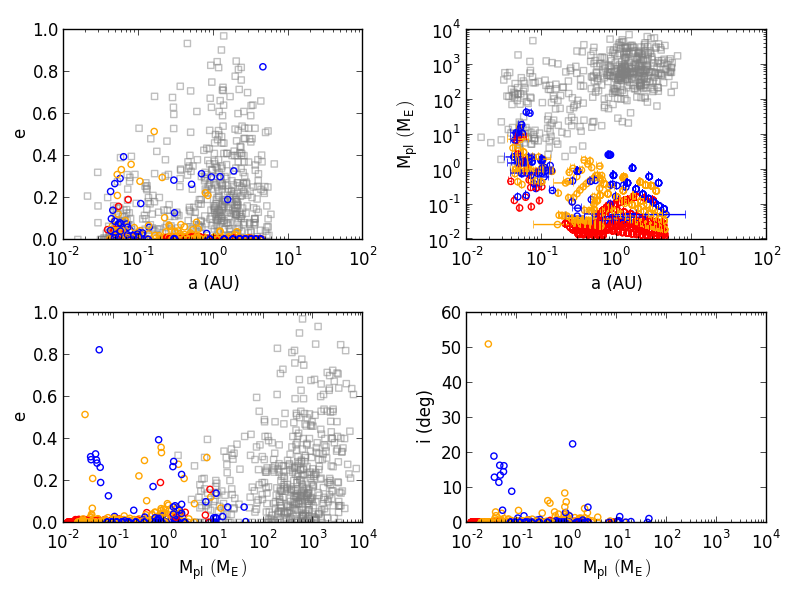}
\caption{The same set of figures as Figure~\ref{fig_aeim_nomig1} for simulations with migration, 
at the end of the simulations (50\,Myr).
\label{fig_aeim_mig}}
\end{figure*}
Figure~\ref{fig_aeim_mig} shows the corresponding results at $50\,$Myr for cases with migration. 
The figure at $4\,$Myr looks very similar to this case and is not shown here.
The $a-M_{pl}$ distribution of Figure~\ref{fig_aeim_mig} looks very different from observations, 
with only a few hot/warm giant planets, no cold giant planets, 
and little variation of semimajor axes especially for SEs. 
The most massive giant planets survived here have $\sim0.17\,M_J$. 
The lack of more massive hot Jupiters (HJs) is due to the too efficient type I migration resulting from 
the choice of our disc model (see Section~\ref{discussion}).
The efficient migration also leads to a cluster of planets around the disc's inner edge. 
Here, we also find that Es/SEs form and migrate to the inner disc region by $\lesssim 1\,$Myr, 
before less massive ones ($\lesssim 1 M_E$) do.
We will discuss the migration issue further in Section~\ref{discussion}.

The eccentricities of planets are overall lower than for simulations without migration. 
This is partly because, in these simulations, 
most dynamical instabilities happen during the migration phase (see Section~\ref{sec_ejemer}) 
and the number of planets per system after the gas disc dissipation tends to be smaller than 
those with no migration.

\begin{figure}[ht]
\includegraphics[width=1.\linewidth]{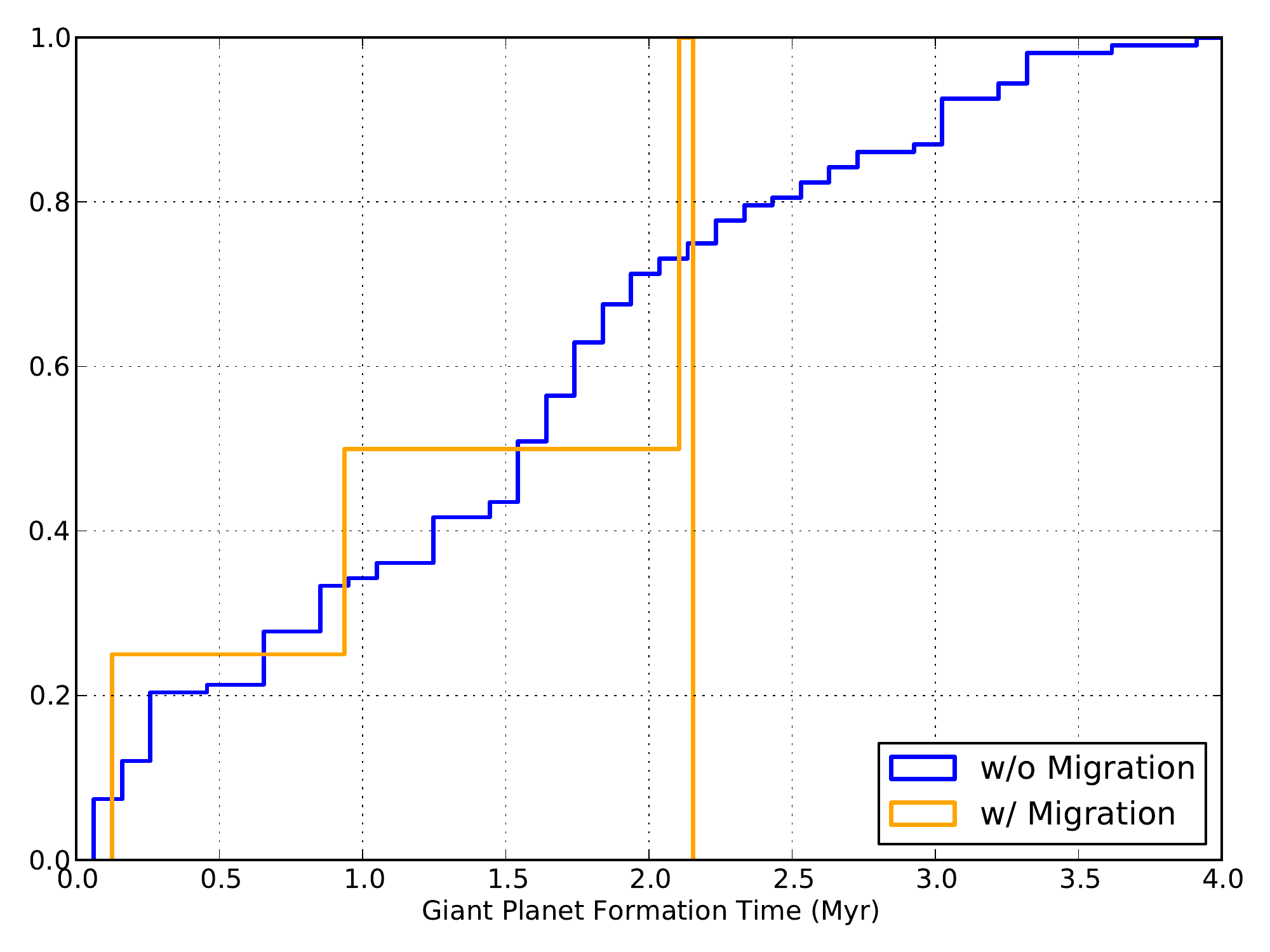}
\caption{The cumulative plot of formation timescales of a giant planet with mass $\geq 0.1\,M_J$ for 
without (blue) and with migration cases (orange). 
About 50\% of giant planets are formed within 1.5\,Myr. \label{fig_tgiant}}
\end{figure}
Figure~\ref{fig_tgiant} shows the cumulative plot of formation time scales 
for all the giant planets ($\geq 0.1\,M_J$) formed in our simulations without (blue) and with migration (orange). 
There are only a few giant planets formed in simulations with migration, but the distribution has a similar trend to 
those without migration. 
We find that half the giant planets in our simulations are formed within $\sim1.5\,$Myr and 
$\sim90\%$ are formed within $\sim3\,$Myr. 
Thus, giant planet formation with pebble accretion is faster than in the classical formation scenario 
and consistent with recent estimates of Jupiter's formation time scale \citep{Kruijer17}.

\begin{figure*}[ht]
\begin{subfigure}{.5\textwidth}
\includegraphics[width=1.\linewidth]{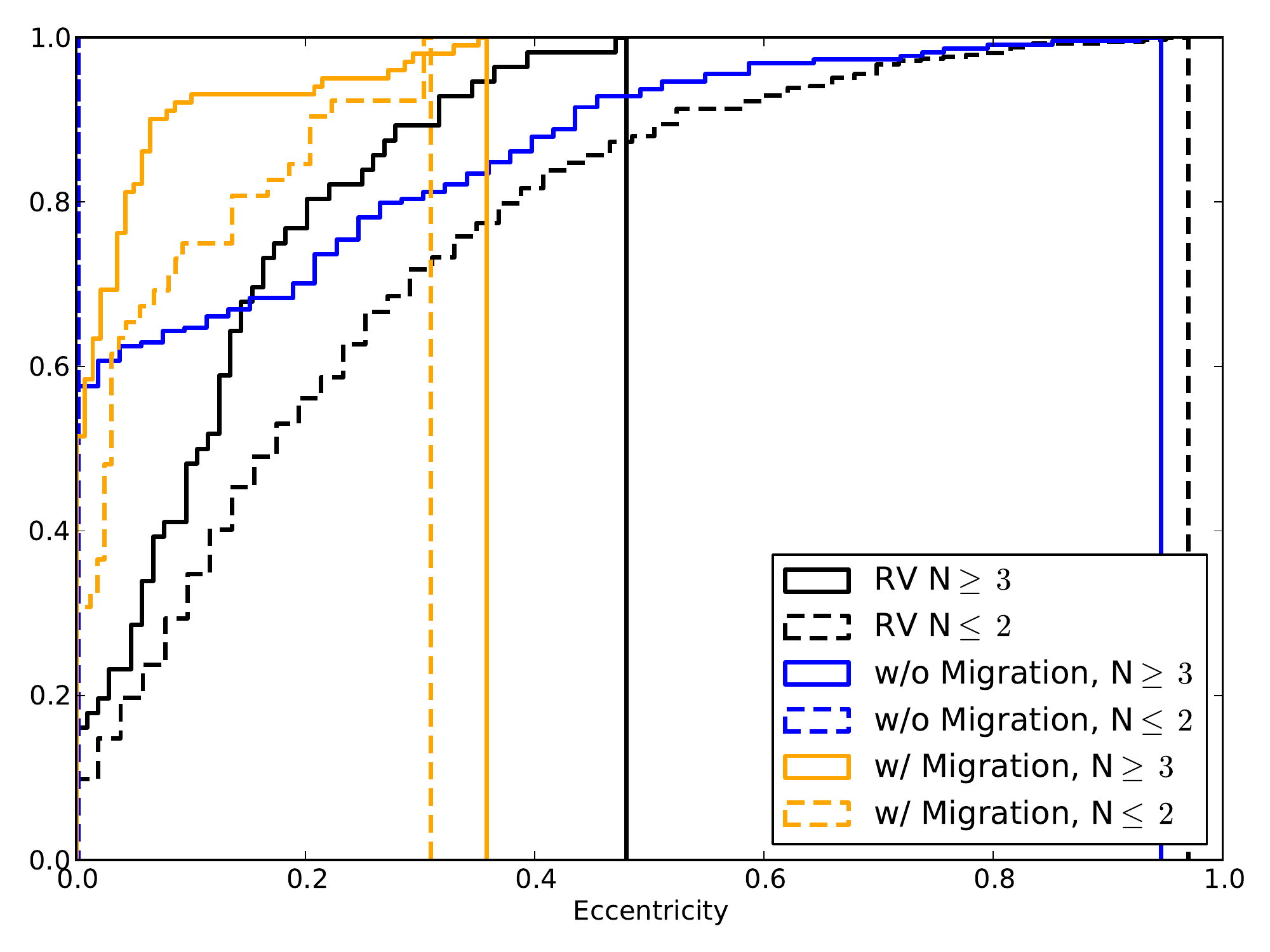}
\end{subfigure}
\begin{subfigure}{.5\textwidth}
\includegraphics[width=1.\linewidth]{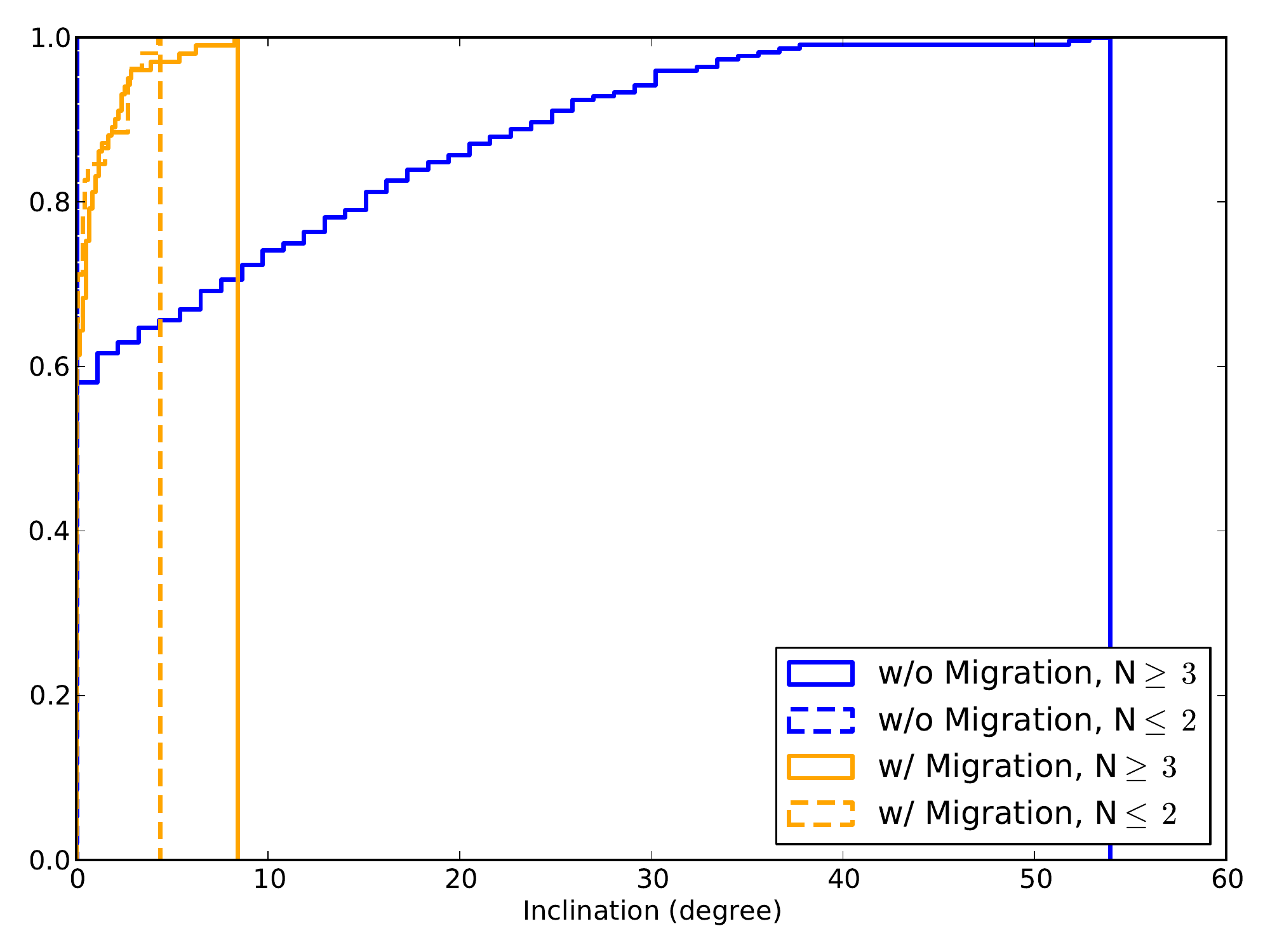}
\end{subfigure}
\caption{Left: The cumulative eccentricity distributions for planets formed in simulations without (blue) and with migration (orange) 
are compared with distributions of planets detected by the RV method (black).   
The dashed and solid lines correspond to systems with one or two planets and more than three planets, respectively.  
Right: The corresponding cumulative distributions for inclinations.
Simulated planets, in particular those from with-migration runs, tend to have nearly circular and coplanar orbits. 
\label{fig_eccinc}}
\end{figure*}
%
Among the observed planets, the maximum eccentricity has been seen to increase with planetary mass 
(see, e.g., grey squares in Figure~\ref{fig_aeim_nomig1}), 
but the eccentricity also decreases with the number of planets per system.
\cite{Limbach15} studied catalogued RV systems and found that the median eccentricity of planetary systems 
decreases with the number of planets $N$ as $e\sim0.584\,N^{-1.20}$ for $N\geq3$, while the median 
eccentricities of systems with $N\leq2$ are comparable.   
Figure~\ref{fig_eccinc} compares the cumulative distributions of planetary eccentricities and inclinations 
for our simulations without 
(blue) and with migration (orange), along with the distributions for observed planetary systems (black).
The dashed and solid lines correspond to systems with one or two planets ($N\leq2$) and 
those with more than three planets ($N\geq3$), respectively.
In both of our simulations, more than $60\%$ of planets have eccentricities $e<0.1$ and 
inclinations $i\sim0^{\circ}$.  
Although these are low even compared to systems with $N\geq3$ planets, our simulations demonstrate that 
eccentricities and inclinations tend to be lower for higher planet multiplicity systems.  
Thus, although the actual distributions are different, their trends are consistent with observations.

\subsection{Formation Efficiency and Stellar Metallicity}\label{sec_met}
\begin{figure*}[ht]
\includegraphics[width=1.\linewidth]{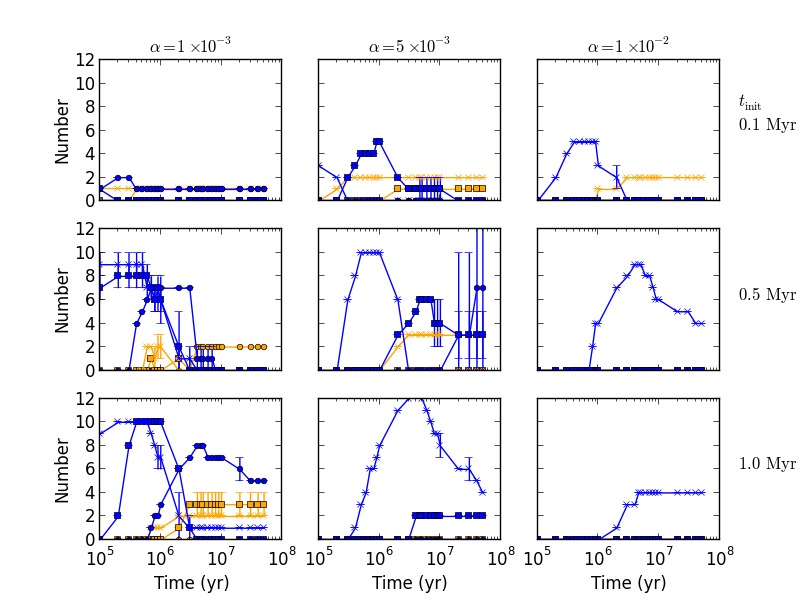}
\caption{Evolution of the average number of planets per system for different combinations of $\alpha$ and $t_{\rm init}$ in 
simulations without migration. 
The orange and blue lines show the giant planets ($\geq 0.1\,M_J\sim30\,M_E$) and Earths or Super-Earths ($1.0\,M_E-0.1\,M_J$), 
respectively.
The cross, square, and circle correspond to the stellar metallicities of ${\rm [Fe/H]}=0.5$, 0.0, and -0.5, respectively. 
Giant planets do not form in high-viscosity and low-mass discs.  
The formation efficiency of all types of planets show some dependency on the stellar metallicity (see text).      
\label{fig_number}}
\end{figure*}
\begin{figure*}[ht]
\includegraphics[width=1.\linewidth]{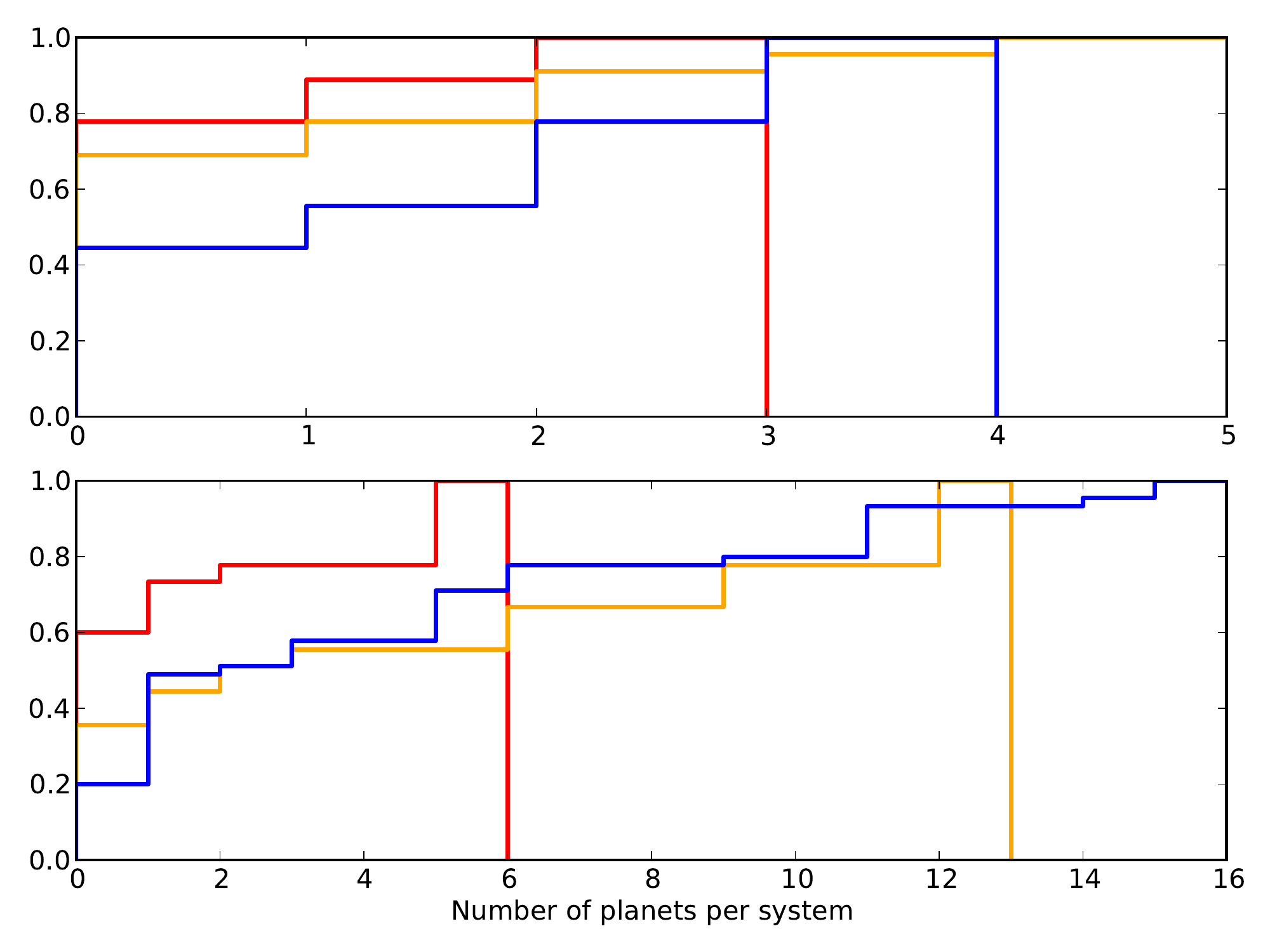}
\caption{The cumulative distributions of a number of planets per system for giant planets ($\geq 0.1\,M_J$, top panel), 
and Es/SEs ($0.1\,M_E-0.1\,M_J$, bottom panel). 
The blue, orange, and red correspond to the stellar metallicities of ${\rm [Fe/H]}=0.5$, 0.0, and -0.5, respectively. 
The fractions of planetary systems with zero planets decrease with stellar metallicities both for giant planets and Es/SEs.     
\label{fig_number2}}
\end{figure*}
In this subsection, we discuss the formation efficiency of planets in terms of a stellar metallicity.  
We focus on no-migration cases here, but simulations with migration have a similar trend.

Figure~\ref{fig_number} shows the evolution of the number of planets per system averaged over five simulations each 
for no migration cases. 
The orange and blue lines represent giant planets ($\geq 0.1\,M_J\sim30\,M_E$) and Es/SEs ($1.0\,M_E-0.1\,M_J$), respectively.
The cross, square, and circle correspond to the stellar metallicities of ${\rm [Fe/H]}=0.5$, 0.0, and -0.5, respectively.

In our simulations, giant planets are formed only for the upper left parameter combinations, where 
a disc is initially more massive and the disc's viscosity is lower.  
The trend is consistent with the previous work with the classical planet formation scenario \citep[e.g.,][]{Thommes08,Coleman16b}. 
In these studies, the lack of giant planets for higher viscosity discs has been attributed to the shorter disc lifetimes.  
Our disc evolution represented by $\dot{M}_*$ is not explicitly associated with the viscosity parameter as seen in Eq.~\ref{eq_M*}.
However, the effect of the disc evolution is implicitly taken into account 
in calculating pebble accretion rate via the pebble mass flux $\dot{M}_{F}$ as in Eq.~\ref{eq_MFdot} 
(i.e., the pebble flux is smaller and thus the accretion rate is lower for a more viscous disc). 

The figure also confirms the well-known dependency of the existence of giant planets on stellar metallicities 
\citep[e.g.,][]{Gonzalez97,Fischer05,Johnson10} --- 
in our simulations, giant planets are more easily formed around higher metallicity stars.
In fact, for the lowest metallicity cases of ${\rm [Fe/H]}=-0.5$, giant planets are only formed 
for $\alpha=10^{-3}$ cases with $t_{\rm init}=0.1$ and 0.5\,Myr, while 
for the highest metallicity cases of ${\rm [Fe/H]}=0.5$, giant planets are formed in 
all cases but $(\alpha, t_{\rm init})=(5\times10^{-3}, \, 1.0\,{\rm Myr})$, 
$(10^{-2}, \, 0.5\,{\rm Myr})$, and $(10^{-2}, \, 1.0\,{\rm Myr})$.

Compared to giant planets, Es and SEs are more easily formed across all combinations of 
$\alpha$ and $t_{\rm init}$.
This is consistent with observations showing that 
Es and SEs are more common than giant planets \citep{Howard10,Mayor11ap,Winn15}.

Our simulations also show the dependence of formation rates of Es/SEs on stellar metallicity.
This is the easiest to see in the lower left corner of Figure~\ref{fig_number}. 
For the higher metallicity environments, 
the formation of Es/SEs is faster and a larger number of them is formed as well.  
In some cases, the average number of Es/SEs reach $>10$. 
The number of planets, however, decreases over time due to planet-planet interactions. 
The final number of Es/SEs in the system depends on the timing of planet formation as well as the later 
dynamical evolution of planets, and thus the dependency of Es/SEs on stellar metallicities is partly washed out. 

These trends are further confirmed in Figure~\ref{fig_number2} which plots the final number of planets per system for 
giant planets ($\geq 0.1\,M_J$, top panel), and Es/SEs ($0.1\,M_E-0.1\,M_J$, bottom panel). 
A fraction of systems with no giant planets decreases roughly from 0.8, 0.7, and 0.4 for stellar metallicities of 
${\rm [Fe/H]}=-0.5$, 0.0, and 0.5. 
Similarly, a fraction of systems with no Es/SEs decreases roughly from 0.6, 0.4, and 0.2 with stellar metallicities. 
The overall number distributions of Es/SEs look similar for ${\rm[Fe/H]}=0.0$ and 0.5, while 
the lowest metallicity cases appear to produce less Es/SEs compare to them. 
However, if we limit the planetary masses to $1.0\,M_E-0.1\,M_J$, the maximum number of planets per system 
is 5 or 6 (also see Figure~\ref{fig_number}) and the distributions for ${\rm[Fe/H]}=-0.5$ and 0.0 are indistinguishable, 
while the highest metallicity cases tend to produce more planets than the others. 
Although the metallicity dependence for Es/SEs is subtle, 
there is an overall trend that higher metallicities lead to a higher number of more massive planets.  

It has been considered that there is no planet-metallicity correlation for planets smaller than giant planets 
\citep[e.g.,][]{Sousa08,Mayor11ap,Neves13}.
However, a recent study showed that a planet-metallicity correlation is universal \citep{Wang15} --- 
not only gas giants, but SEs/Es occur more frequently around metal-rich stars.
The suggestion has been debated \citep{Buchhave15}, but our study shows that there may be a metallicity 
dependence even for Es/SEs.

\subsection{Mass Fractions of Cores and Envelopes}\label{sec_composition}
\begin{figure*}[ht]
\begin{subfigure}{.5\textwidth}
\includegraphics[width=1.\linewidth]{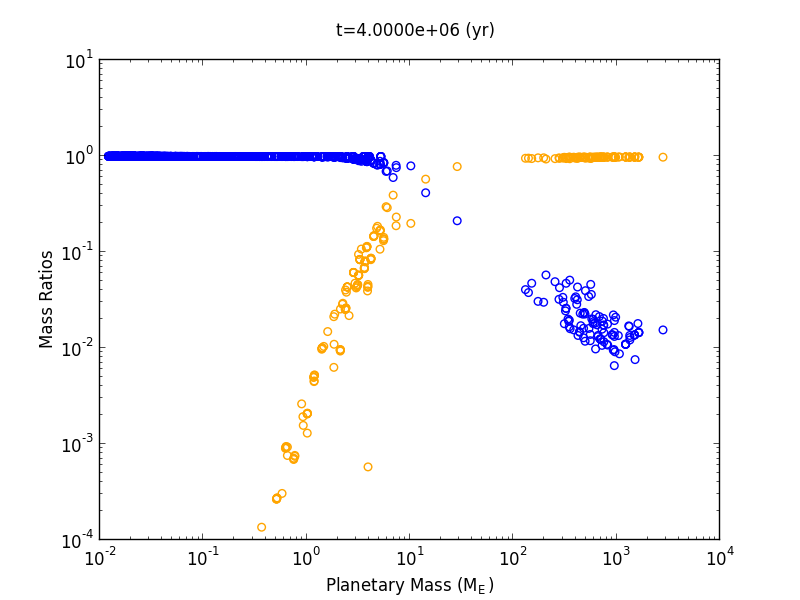}
\end{subfigure}
\begin{subfigure}{.5\textwidth}
\includegraphics[width=1.\linewidth]{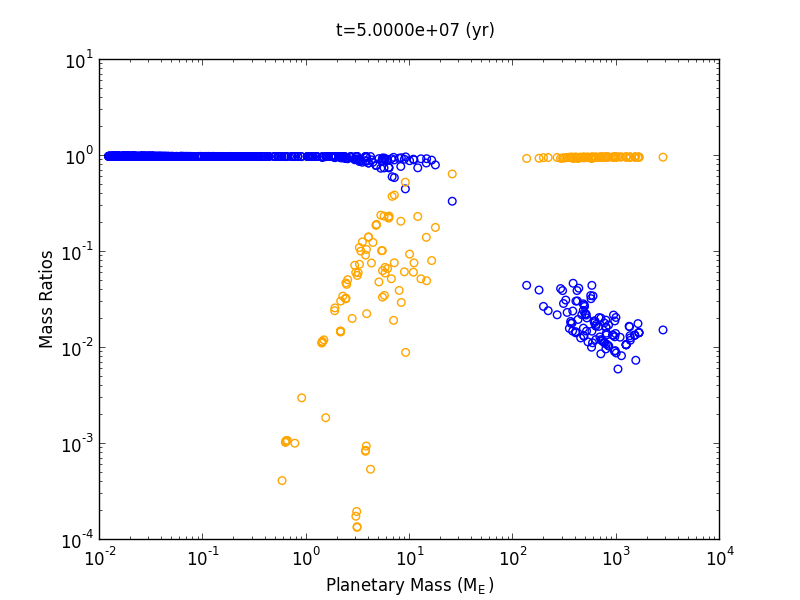}
\end{subfigure}
\caption{The core and envelope mass fractions for different planetary masses just before the disc dissipation (at 4\,Myr, left panel), 
and at the end of the simulations (at 50\,Myr, right panel). 
Orange and blue circles represent mass fractions of cores $M_{\rm core}/M_pl$ and envelopes $M_{\rm env}/M_pl$, respectively. 
Mass ratios for the core and envelope change significantly for Es/SEs due to the planet-planet collisions after the disc dissipation. 
\label{fig_massratios}}
\end{figure*}
%
Recent observations have revealed that Es/SEs come in a wide range of densities \citep[e.g.,][]{Marcy14,Weiss14}. 
From the planet models, the ranges of densities indicate that some planets have nearly no envelopes while 
others have several tens of \% of mass in envelopes \citep[e.g.,][]{Howe14,Lopez14}.  
Figure~\ref{fig_massratios} compares the mass fractions of core and envelope masses 
($M_{\rm core}/M_{pl}$ and $M_{\rm env}/M_{pl}$ for orange and blue circles, respectively) for all planets formed 
in the no-migration simulations. 
Just before the disc's dissipation (4\,Myr), Earth-mass planets have $\lesssim 1\%$ masses in the envelope, 
$\sim10M_E$ SEs have roughly equal amount of mass in the core and envelope, and Jupiter-mass planets have a few 
to several \% of masses in the core.  
This picture changes by the end of the simulations (50\,Myr), especially for Es/SEs.  
Due to planet-planet collisions, we see that SEs could have a variety of densities, 
from less than $1\%$ of mass in the envelope to roughly equal amount of mass in the envelope and the core.
Our results show that the dynamical evolution alone already predicts the diversity in densities. 

The corresponding results for simulations with migration look similar.  
However, since the dynamical instability sets in earlier in these cases (see Section~\ref{sec_ejemer}), 
the variety of planets emerges while the gas disc is still around.

\subsection{Period Ratios of Planets}\label{sec_pratios}
\begin{figure*}[ht]
\includegraphics[width=1.\linewidth]{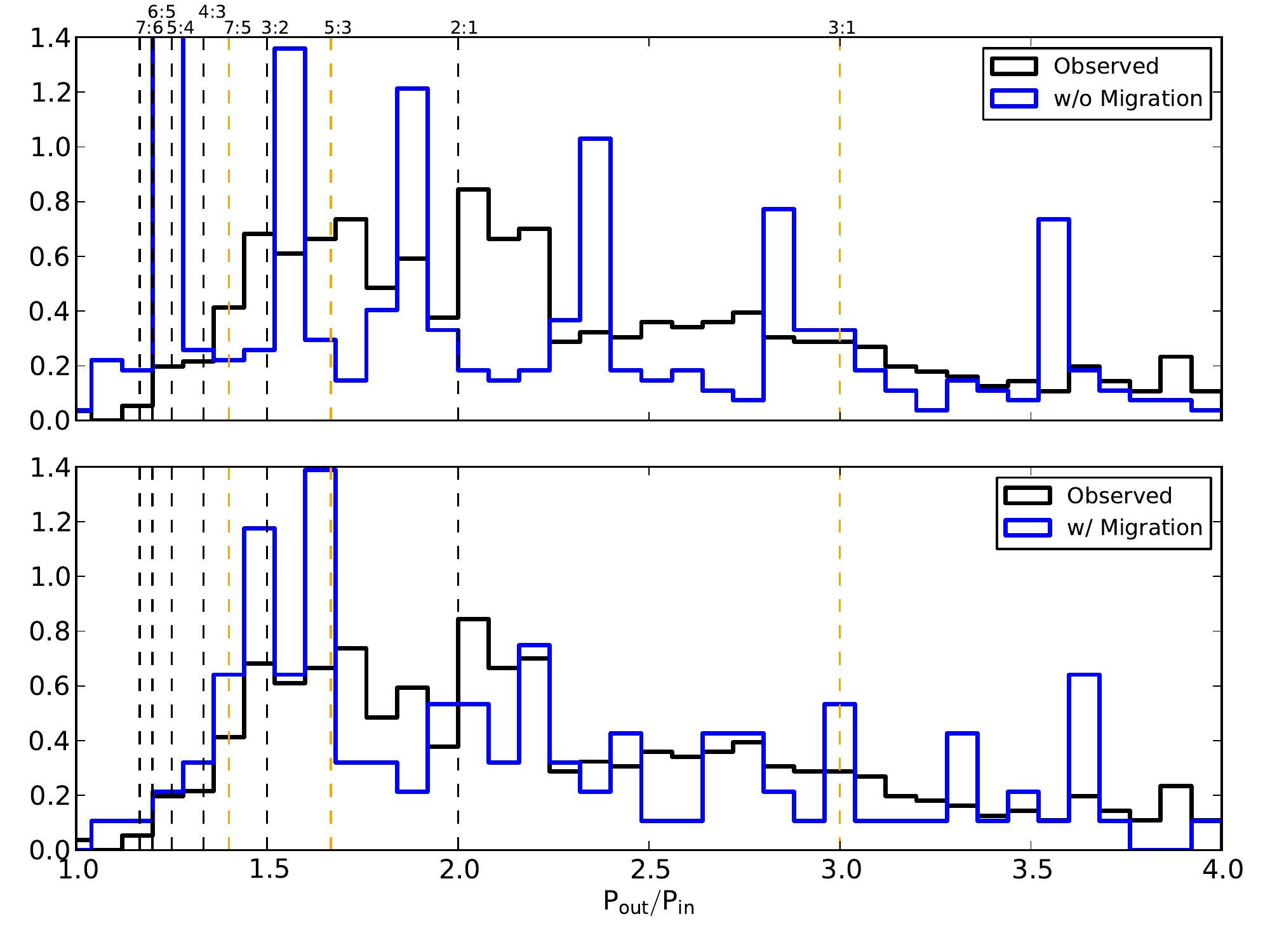}
\caption{The distributions of period ratios of all planet pairs (not just adjacent ones) in multiple-planet systems 
which have only planets with masses above $0.1\,M_E$.  
The top and bottom panels show these distributions for planets formed in simulations without and with migration, respectively. 
The corresponding values for confirmed planets that are observed by all kinds of detection methods are shown in black. 
The peak near the 6:5 resonance is due to initial separations of planetary embryos (see text for details).  
The planetary systems formed in our simulations tend to be more compact than typical observed planetary systems, 
but the trend agrees well with that for Kepler planets \citep{Fabrycky14}. 
\label{fig_pratios}}
\end{figure*}
\cite{Fabrycky14} studied Kepler's multiple-planet systems and showed that the period ratios of all planet pairs 
(not just adjacent ones) have the mode of the distribution slightly wide of the 3:2 resonance and that 
there are few planet pairs interior to the 5:4 resonance.
Figure~\ref{fig_pratios} shows such period ratios for all planet pairs in a multi-planet system from 
simulations without (top panel) and with migration (bottom panel).
We only include systems where planetary formation proceeds sufficiently so that all planets have masses $\geq 0.1\,M_E$.
These distributions are compared with a corresponding distribution for observed planets (black histogram), 
where we use all the confirmed multiple-planet systems (not only RV detected ones).

For the cases with no migration, we find that the planetary separations are clustered around certain period ratios. 
This is partly due to the fact that our initial separation of seed planets is near the 6:5 resonance.  
Thus, the peaks near 1.2, 2.4, and 3.6 are explained by considering period ratios of planet pairs in relatively dynamically 
stable systems.  
By plotting period ratios for only neighbouring pairs, we find that the peaks near the 3:2, 2:1 and 3:1 resonances are much 
less pronounced, with the one slightly wide of the 2:1 being most prominent.  

For the cases with migration, there is no such artificial peak.  
The distribution has a good agreement with what is described in \cite{Fabrycky14}, with the peak near the 3:2 resonance 
and the decline in a number of planet pairs toward the 5:4 resonance and shorter separations. 
The distribution also agrees well with the period ratio distribution of observed pairs from all detection methods near the 
short separation end.  
However, the agreement between the 3:2 to 2:1 resonances is not very good, because 
the observed distribution has a broad peak in this region while the simulated result has a peak near the 3:2 and 5:3 resonances. 
This indicates that planetary systems in our simulations are more compact than the average observed planetary systems and 
are closer to Kepler's multiple-planet systems. 
The tendency for compact systems is consistent with the preference for low eccentricities and inclinations 
of our planets seen in Figures~\ref{fig_aeim_nomig2} and \ref{fig_aeim_mig}.
%
%
%
%
\subsection{Dynamical Instability and Its Outcomes}\label{sec_ejemer}
\begin{figure*}[ht]
\includegraphics[width=1.\linewidth]{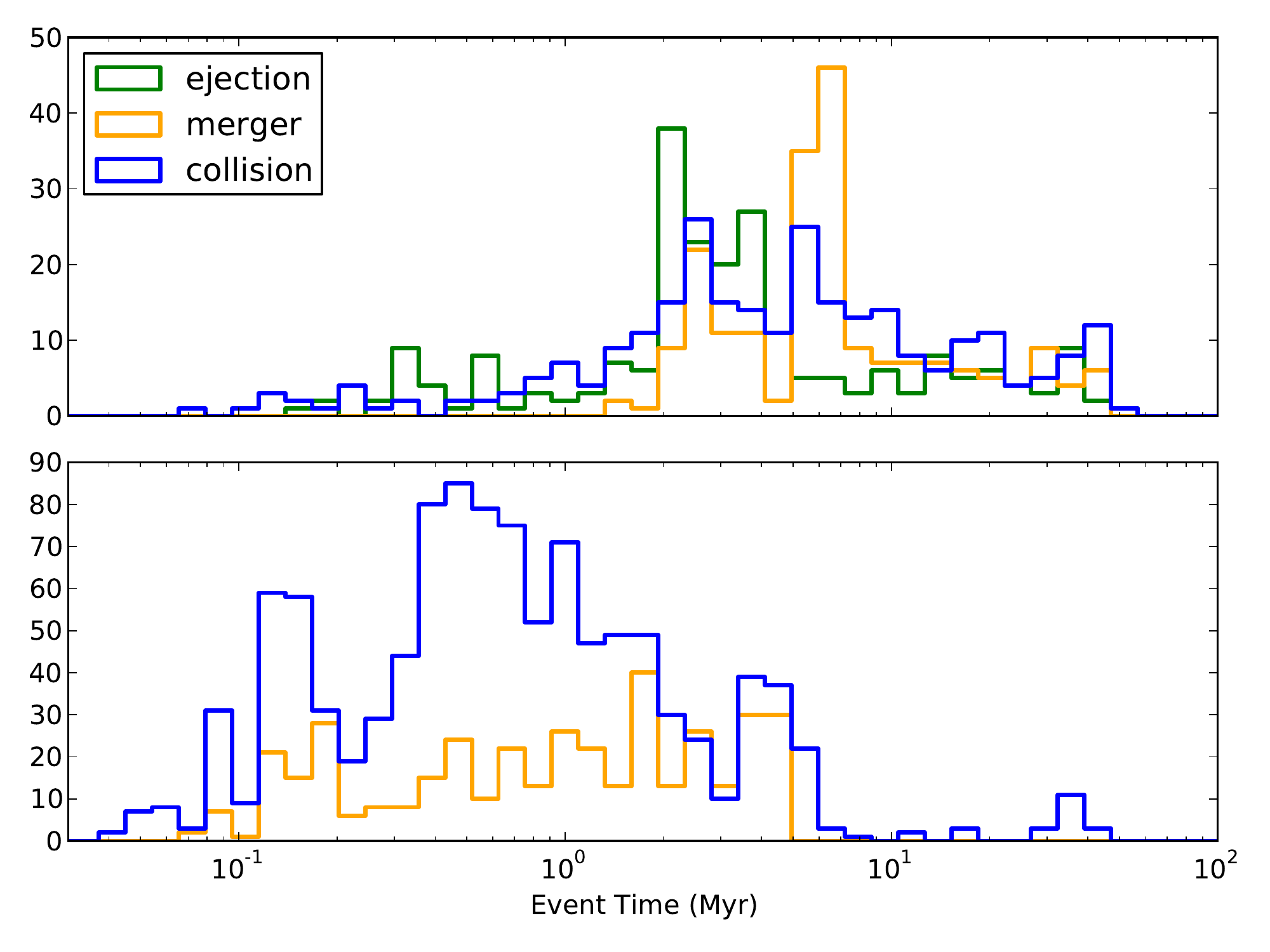}
\caption{The distributions of the timing of planet-removing events for without (top) and with migration (bottom). 
Green, orange, and blue histograms correspond to ejection of a planet from the system, merger of a planet with the central star, and 
planet-planet collision, respectively. 
Dynamical instability tends to occur as the disc disappears for runs with no migration, and before the disc dissipation 
for runs with migration. 
\label{fig_cem}}
\end{figure*}
%
%
As seen in Figures~\ref{fig_aeim_nomig2} and \ref{fig_aeim_mig}, orbital eccentricities and inclinations for runs with migration 
tend to be lower than those for runs without migration. 
This is largely due to the timing of dynamical evolution. 
Figure~\ref{fig_cem} shows when planet's removal events, such as ejection, planet-planet collision, and planet-star merger, occur 
in our simulations.  
For the cases without migration, most dynamical instability events occur while the gas disc is around ($\lesssim4\,$Myr), 
while for the cases with migration, the dynamical instability sets in as the disc dissipates. 
As a result, planetary eccentricities and inclinations in the former case could be damped after the dynamical instability 
by the still-existing gas disc.  

\begin{figure*}[ht]
\begin{subfigure}{.5\textwidth}
\includegraphics[width=1.\linewidth]{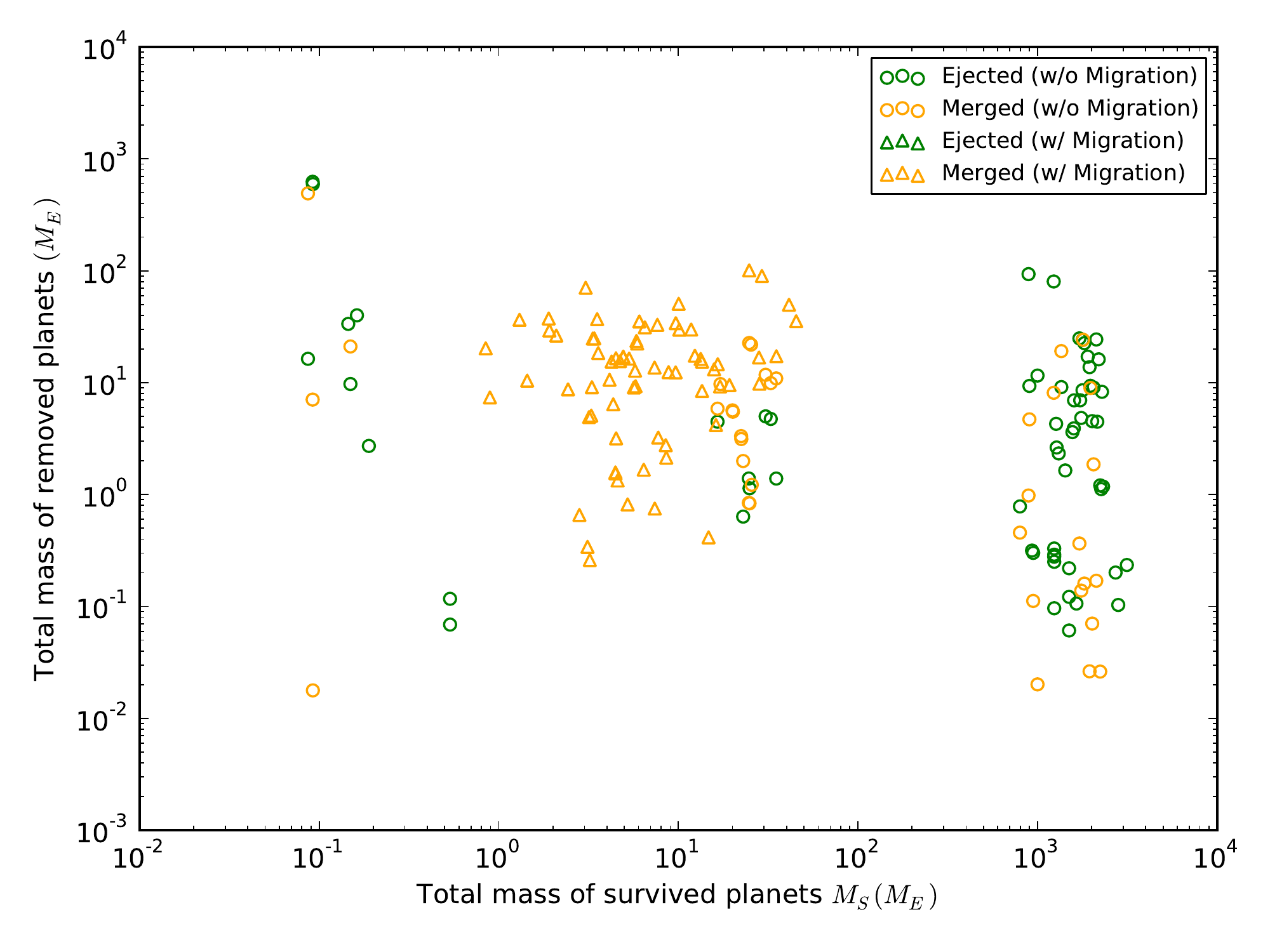}
\end{subfigure}
\begin{subfigure}{.5\textwidth}
\includegraphics[width=1.\linewidth]{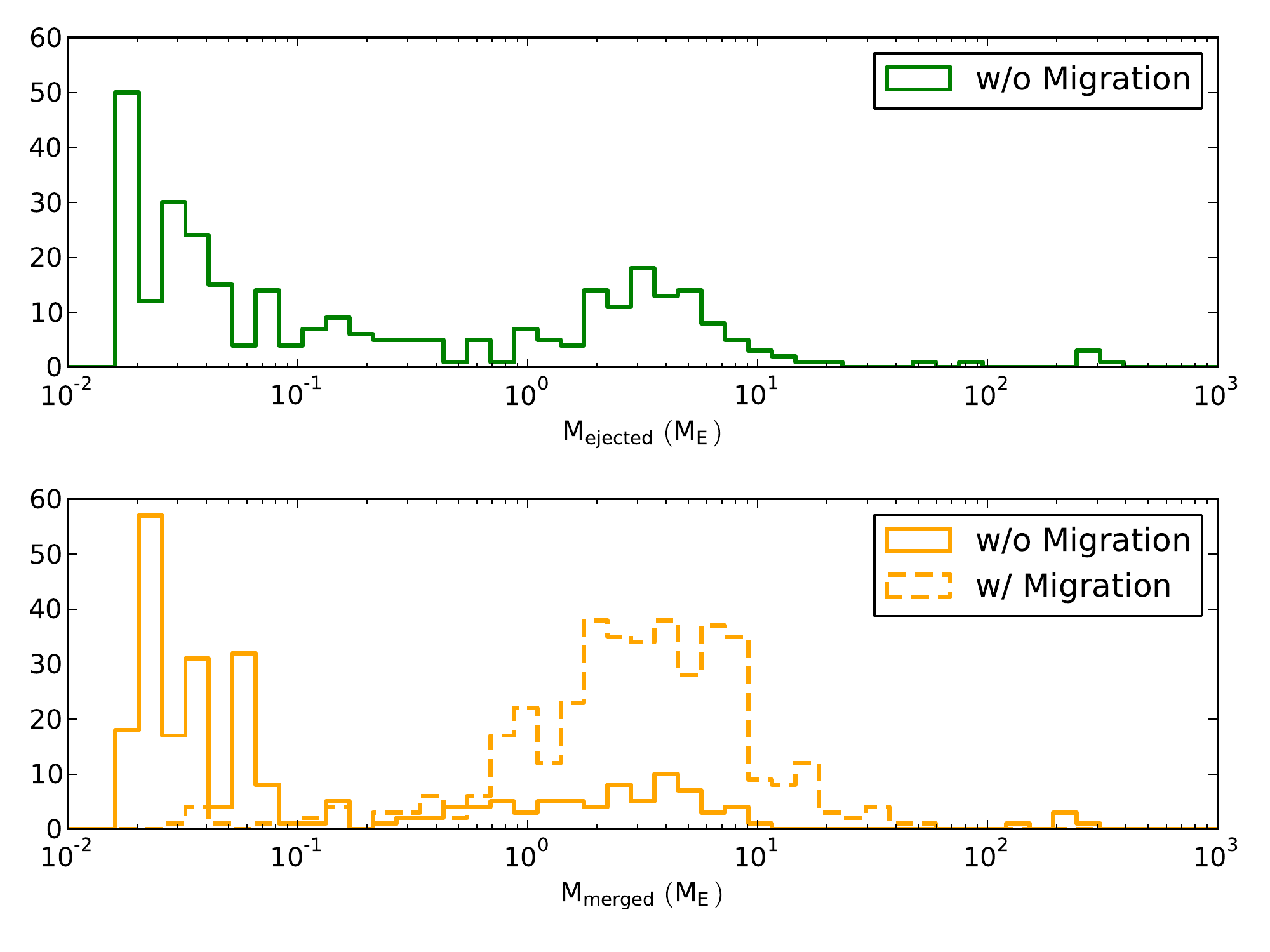}
\end{subfigure}
\caption{Left: The comparison of the total mass in removed planets with that in survived planets. 
Green and orange symbols correspond to ejection of a planet from the system and merger of a planet with the central star, respectively. 
Also, circles and triangles are for runs withut and with migration, respectively.
There is no correlation between the ejected/merged masses and the masses in survived planets.
Right: The mass distributions of ejected (top panel) and merged planets (bottom panel). 
The outcome of with-migration cases is plotted in dashed line.
\label{fig_memass}}
\end{figure*}
The left panel of Figure~\ref{fig_memass} compares the masses ejected from the systems (green) 
and those merged with the central star (orange) in terms of the total mass in survived planets. 
Circle and triangle symbols are without and with migration cases, respectively. 
As also shown in Figure~\ref{fig_cem}, there are no ejections in with-migration cases, 
but the ejection and merger events occur at nearly equal frequency for no-migration cases.
We find no obvious dependence of the total masses of removed planets on those of survived ones. 
The removed masses per system could range from $0.01-10^3\,M_E$ with median values of ejected and 
merged masses being $4.3$ and $9.8\,M_E$, respectively. 

The right panel of Figure~\ref{fig_memass} shows the distributions of planetary masses ejected from the system or merged 
with the central star. 
For no-migration simulations, the production rate of free-floating giant planets ($\geq0.1\,M_J$) is low 
($\sim 1.94\%$ of all ejected planets), but the free-floating Es/SEs ($0.1\,M_E-0.1\,M_J$) is 
significant ($\sim 48.9\%$ of all ejected planets). 
The distribution is similar for merger rates: merging giant planets and Es/SEs are $\sim 1.98\%$ and $\sim31.3\%$ of 
all merged planets. 
The trend of the distribution, however, is significantly different for with-migration simulations.  
Most merged planets are Es/SEs ($\sim96.7\%$) while $\sim1.27\%$ of merged planets are giant planets. 

A recent work by \cite{Barclay17ap} studied the late stage ($\gtrsim10\,$Myr) of terrestrial planet formation 
out of Mars-mass embryos over $0.3-4.0\,$AU with and without giant planets, and showed that 
(1) there was little ejection without giant planets, and 
(2) although the individual mass of half of the ejected planets was $>0.06\,M_E$, there were no planets more massive than $0.3\,M_E$. 
They further suggested that WFIRST would discover up to 20 Mars-mass planets but few free-floating Earth-mass planets. 
Our simulations support their results partly, because the median mass of ejected planets is about Mars mass ($0.11\,M_E$) 
--- a significant fraction of ejected planets are of low mass.
However, our work also suggests that (1) the existence of giant planets is not a necessary condition for generating 
free-floating planets, and (2) it is possible to have Es/SEs as free-floating planets. 
In fact, in our simulations, the ratio of ejected planets with $1.0\,M_E-0.1\,M_J$ and those with less than $1.0\,M_E$ is 0.5. 
Therefore, we would expect at least one E/SE free-floating planet per two Mars-like ones.  
The results, however, should be taken with caution because our simulations with migration do not lead to any ejections 
due to early dynamical instability and our simulations do not reproduce the observed distributions well. 
%
%
%
%
\section{Discussion}\label{discussion}
\subsection{Eccentricity and Planetary Mass}\label{sec_eM}
For simulations without migration, it has not been expected that the observed semimajor axis distribution would be recovered. 
On the other hand, the observed eccentricity distribution could have been reproduced if planetary systems that formed in our simulations 
represented a typical set of exoplanetary systems, because 
the eccentricity distribution is largely determined by planet-planet interactions \citep[e.g.,][]{Ford08,Chatterjee08,Juric08}. 
Figures~\ref{fig_aeim_nomig2}, \ref{fig_aeim_mig}, and \ref{fig_eccinc} demonstrate that the planets formed in our simulations 
tend to have too low eccentricities (and inclinations).  
More specifically, all planets with high eccentricities ($e>0.1$) in our simulations are SEs or smaller ($< 30M_E$) and 
all giant planets have low eccentricities.  
In Section~\ref{sec_aMall}, we have argued that the lack of eccentric giant planets is likely due to a small number of 
giant planets per system (see Figure~\ref{fig_number}).

\cite{Ida13} presented a comprehensive population synthesis model where they took account of the effects of close encounters between 
planets by calibrating the analytical model with N-body simulations. 
Their simulations showed a very good agreement with observations especially for $a-e$ and $M_{pl}-e$ distributions, and 
reproduced the trends that the eccentricity of observable planets 
(i.e., the stellar RV of $\geq1\,$m/s and orbital periods of $\leq10\,$yr) increased with semimajor axis and planetary mass.
They also found that highly eccentric giant planets were often the remnants of massive discs where several giant planets were formed, 
while moderate disc masses (comparable to the minimum mass solar nebula model) commonly 
led to one or two relatively small-mass planets with low eccentricities.  
The latter is similar to typical systems with giant planets in our simulations.

This may seem to imply that we need to consider a wider range of disc masses than we use here to reproduce the eccentricity 
distribution. 
However, our initial disc masses that are relevant for planet formation are comparable to theirs. 
The direct comparison of disc masses for classical planet formation and pebble accretion is difficult, 
because planetesimal-based planet formation is largely limited to $\lesssim30\,$AU while the entire disc (especially the 
outer one) is important for pebble accretion. 
Here, we estimate the initial disc masses in \cite{Ida13} as $1.6\times10^{-4} - 0.16\,M_{\odot}$ by assuming 
the Sun-like star with a disc size of $0.1-30\,$AU.
On the other hand, the initial disc masses in our simulations with a disc size of $0.1-100\,$AU 
range over $7.4\times10^{-3} - 0.39\,M_{\odot}$, by ignoring the unrealistically high disc mass arising 
from the combination of $\alpha_3=1$ and $t_{\rm init}=0.1\,$Myr.
Thus, the disc masses relevant for planet formation in this work are comparable to (or even higher than) 
those in \cite{Ida13}. 

The difference seems to be how the gas disc decays.
In \cite{Ida13}, the disc mass decays exponentially with the disc's lifetime of $1-10\,$Myr, while in our simulations, the 
disc mass decreases as the stellar mass accretion rate decreases as in Eq.~\ref{eq_M*} and the disc lifetime is $4-5\,$Myr.  
Therefore, the most massive and long-lived disc in \cite{Ida13} would have $0.14\,M_{\odot}$ after $1\,$Myr while 
in all of our simulations, the disc mass is below $0.074\,M_{\odot}$ by $1\,$Myr.   
Our future study will take account of not only a range of disc masses, but also how the gas disc decays.
\subsection{Migration and Survival of Planets}\label{sec_mig}
For simulations with migration, we have encountered a problem of retaining planets.
As seen in Figure~\ref{fig_aeim_mig} of Section~\ref{sec_aMall}, most planets in our simulations 
have been lost to the central star despite that we took account of non-isothermal effects of the disc, and 
also made a ``trap'' at the disc's inner edge (see Section~\ref{discmodel}). 
In this subsection, we discuss both of these effects further.

\subsubsection{Effects of type I migration}\label{sec_typeI}
The overly efficient type I migration resulted in a cluster of Es/SEs-like planets around the disc edge with a small variation 
in semimajor axes and a lack of massive giant planets (see the $a-M_{pl}$ distribution of Figure~\ref{fig_aeim_mig}). 
The trend is very different from observations, but similar to the previous work done by \cite{Coleman14}.
They studied classical planet formation (i.e., planetesimal accretion rather than pebble accretion) 
in thermally evolving viscous disc models by using N-body simulations 
and found that, for giant planets to form and survive beyond the disc's inner edge, runaway gas accretion and the transition 
from type I to type II migration needed to occur while planets were still far from the central star.

\cite{Ida13}, on the other hand, successfully retained more Es/SEs as well as hot Jupiters in their population synthesis model 
of classical planet formation (see their Figure~14).
This indicates that the outcome of non-isothermal type I migration sensitively depends on a disc model.
However, their model also failed to reproduce the observed overdensity of gas giants beyond $\sim0.7\,$AU 
and their distribution predicts an under-population of planets with $10-10^2\,M_E$ within a few AU, which is not 
clear from the observation.

This suggests that there may be a fundamental mechanism that we are missing in reproducing the $a-M_{pl}$ distribution.
\cite{Coleman16b}, for example, resolved the migration issue in \cite{Coleman14} by considering transient radial structures of a disc 
(i.e., planet traps), 
where the viscous stress is temporally varying over a range of disc radii.  
Although the number of simulations is not very large, Figure~6 of \cite{Coleman16b} seems to have a good agreement with 
observations, including HJs and cold Jupiters with few giant planets in between.

The problem of a too rapid type I migration was also investigated in \cite{Brasser17pre}, 
where we studied single-planet formation via pebble accretion in various disc models.   
We showed that the migration directions sensitively depended on the disc structure. 
For a disc model with the temperature gradient of $-9/10$ 
(as in our viscous region, see Eq.~\ref{eq_temp}), most type I migrating planets are lost to the central star 
except for a low-mass planet (less than a few ${\rm M_E}$) in a disc with a large viscosity parameter ($\alpha>4\times10^{-3}$). 
On the other hand, for a disc with a steeper gradient of $-6/5$, most planets may be saved.  
We will consider a more detailed disc model in a future work that also accounts for opacity transitions.

\subsubsection{Effects of the disc's inner edge}\label{sec_inneredge}
\cite{Ogihara10} investigated the effect of the disc's inner edge on stopping planet migration. 
They have shown that it is possible to stop type I migration of a chain of planets, 
when the damping of the eccentricity of the innermost planet by the gas disc overcomes 
the eccentricity pumping by resonantly-trapped outer planets. 

The critical parameters that determine the fate of resonant planets near the disc's inner edge are 
the relative width of the edge, $\Delta r/r_{\rm tr}$ (where $\Delta r$ is the disc edge width), 
and the relative time scales of eccentricity damping and migration, $\tau_e/\tau_a$. 
In our prescription, the latter is
\begin{equation}
 \frac{\tau_e}{\tau_a} = 1.282\,\left|\frac{\Gamma}{\Gamma_0}\right|\,\hat{h}_g^2+e^2 \ ,
\end{equation}
where $\Gamma/\Gamma_0$ is the normalised torque discussed in Section~\ref{migmodel} and we assume $e \ll 1$. 
For nominal parameters of $\hat{h}_g=0.027$ at 0.1~AU, 
$\vert \Gamma/\Gamma_0 \vert \sim 1$ and $e \sim 10^{-3}$, 
we have $\tau_e/\tau_a \sim 0.001$. 
The relative width of the edge we have chosen is $\Delta r/r_{\rm tr}=0.37$. 
Therefore, from Figure~2 of \cite{Ogihara10}, the edge of our disc is likely too wide 
to trap the planets in the resonance. 
If we had chosen $r_{\rm tr} -r_{\rm in} \sim h_g$, then the relative width of the edge is 0.025 
and trapping could be easier. 

\cite{Ogihara10} employed a different migration prescription and 
their planets had a constant mass. 
These differences may have contributed to the contrasting outcomes between our simulations and theirs.
We will investigate this issue further in a future study.

\subsubsection{Type II migration}\label{sec_typeII}
Another potential issue is related to the expression of type II migration.  
The former expression in the bracket in Eq.~\ref{eq_typeII} is the classical type II migration rate 
that is independent of a disc mass and a planetary mass.
Recent hydrodynamic simulations, however, showed that the migration rate of giant planets 
in protoplanetary discs could be faster or slower than this conventional type II migration rate 
\citep[e.g.,][]{Duffell14,Durmann15}, because a gap opened by a planet is not clean and 
gas crossing the gap could contribute to the torque.  
For a Jupiter-mass planet, \cite{Durmann15} found that the migration rate is slower than 
the classical type II migration in a disc with $< 0.2\,M_J$ and faster by 
about a factor of a few in a disc with $\sim M_J$.
We have not taken account of this effect in the current paper, 
but a future work will explore this further. 

\subsection{Gas Accretion}
\subsubsection{Gas accretion near the central star}
In our simulations, the gas accretion rate is determined by Equations~\ref{eq_Mg1} and \ref{eq_Mg2}, independent 
of a distance of planets from the star.
\cite{Ormel15b} studied the properties of atmospheres around low-mass planets via hydrodynamic simulations. 
They showed that the atmosphere of embedded protoplanets replenished on a time scale shorter than 
the Kelvin-Helmholtz time scale in the inner disc and thus hot Jupiters might not be formed in-situ. 

The replenishment time of the atmosphere of a low-mass embedded planet is \citep{Ormel15b}:
%
\begin{eqnarray}
\tau_r &=& \frac{\chi_{\rm env} \hat{h}_g^6 Q}{f_c \Omega}\left(\frac{M_{pl}}{M_*}\right)^{-2} \nonumber \\
       &=& 3.3 \times 10^{11} \frac{\alpha \chi_{\rm env} \hat{h}_g^9 \tau_*}{f_c}
  \left(\frac{M_*}{M_\odot}\right)^2\left(\frac{M_{pl}}{M_E}\right)^{-2} \, {\rm yr} \ , 
\end{eqnarray}
where $Q$ is the Toomre's Q parameter for the gas disc, $\chi_{\rm env}=M_{\rm env}/M_{pl}$ is the mass fraction 
of the envelope, and $f_c\lesssim 1$ is the fraction of streams that reach the shell where the most atmospheric mass resides.
To get the final expression, we also define $\tau_* = M_*/\dot{M}_*$.  
From the expression of $\hat{h}_g$ in Eq.~\ref{eq_hr}, this replenishment time is about a factor of 2 shorter at $0.1\,$AU 
compared to $1\,$AU since $\hat{h}_g$ has a weak dependence on distance.

Equating this to the Kelvin-Helmholtz time scale (Eq.~\ref{eq_tauKH}), 
we can determine the critical mass of an embedded planet above which the envelope gravitationally contracts:
\begin{equation}
\frac{M_{pl}}{M_E} = 0.003\,\frac{f_c}{\alpha\,\chi_{\rm env}\,\hat{h}_g^9\,\tau_*}\Bigl(\frac{M_*}{M_\odot}\Bigr)^{-2} \ .
\end{equation}
With nominal parameters of $\tau_* = 10^8$~yr, $\hat{h}_g=0.027$, $\alpha=10^{-3}$ and $M_*=1 M_\odot$, 
we compute $M_{pl} = 3.9\times10^6\,f_c/\chi_{\rm env} \, M_E$ at 1~AU. 
Unless $f_c/\chi_{\rm env}$ becomes very small, the critical mass is unrealistically large and a giant planet is 
unlikely to be formed at 1~AU. 

The KH time scale, however, could be much shorter when the atmosphere is grain-free \citep{Hori10}. 
\cite{Ormel15b} fitted the calculations by \cite{Hori10} for a grain-free (but not metal-free) atmosphere 
and derived the following time scale. 
%
\begin{equation}
 \tau_{\rm KH}^{\prime} = 10^7\,\chi_{\rm env}^2 \left(\frac{M_{pl}}{M_E}\right)^{-9/2} \, {\rm yr} 
\end{equation}
With this, the corresponding critical mass is written as 
\begin{equation}
\frac{M_{pl}}{M_E} = 0.016\,\hat{h}_g^{-18/5}\left(\frac{\chi_{\rm env} \, f_c}{\alpha \, \tau_*}\right)^{2/5}
\left(\frac{M_*}{M_\odot}\right)^{-4/5} \ ,
\end{equation}
which yields $M_{pl}=68.8\,(\chi_{\rm env}\,f_c)^{2/5} \, M_E$ for nominal parameters. 
The isolation mass at 1~AU is $3.3\,M_E$, and thus the envelope contraction will not happen at 1~AU 
unless $\chi_{\rm env}\,f_c \lesssim 5\times10^{-4}$.
Therefore, even in the grain-free case, giant planet formation is unlikely at 1~AU and 
more difficult within this orbital radius. 
This supports the idea that giant planets formed farther out and then somehow migrated to the current locations to form 
hot Jupiters.

In summary, the critical mass at which the envelope contraction happens depends sensitively on the grainyness of the atmosphere. 
The Kepler mission found a large number of SEs and mini-Neptunes \citep{Fressin13,Petigura13}.
The RV follow-up analysis shows that the density of a substantial fraction of these planets is relatively low 
\citep[e.g.,][]{Weiss14}, indicating the presence of a non-negligible atmosphere. 
\cite{Ormel15b} suggest that the prevention of atmospheric collapse could be responsible for the preponderance of these planets.
Perhaps these planets originally possessed a mass-dominant solid core overlaid by a thin H/He-rich envelope \citep{Lopez14}, 
and they somehow avoided crossing the critical core mass (for which 
$\chi_{\rm env} \sim 1$) during the gas-rich circumstellar disc phase and 
never became hot Jupiters because their atmospheres never contracted. 
The effect of this rapid recycling of atmospheric gas compared to the cooling time scale will be 
considered in a future study.
%
\subsubsection{Critical core mass}
To determine the critical core mass to start gas accretion, we have adopted Eq. (17).  
However, the equation was derived by assuming that the energy from planetesimal accretion was 
deposited at the core-envelope boundary \citep{Ikoma00}. 
It is unclear whether the same equation holds for pebble accretion. 

A recent study by \cite{Alibert17ap} pointed out that the vaporisation of solids in the planetary atmosphere 
could limit the mass of a protoplanetary core, 
if the planetary envelope replenishment timescale was short enough \citep{Ormel15b} and 
the vaporised solids dispersed back into the protoplanetary disc.
By comparing pebble and planetesimal accretion, he found that the maximum embryo mass 
for a typical pebble accretion rate ($10^{-5}\,M_E/{\rm yr}$) is about an Earth mass 
because its envelope would become massive enough to vaporise pebbles, while that for 
a typical planetesimal accretion rate ($10^{-6}\,M_E/{\rm yr}$) could be around ten Earth masses.  
Since such a small planet does not have an efficient gas accretion (Eq.~\ref{eq_tauKH}), it is possible that 
embryos formed solely by pebble accretion never become gas giants.  
The issue should be investigated further in future studies.

\subsection{Formation of close-in low-mass planets}\label{sec_lowmasspl}
\cite{Fulton17ap} studied close-in (orbital periods of $< 100\,$day) small Kepler planets and 
found that the distribution of their radii are bimodal, 
with peaks at 1.3 and $2.4\,R_E$ and a paucity of planets between $1.5-2.0\,R_E$. 
The existence of this paucity was predicted to result from the photoevaporation 
of planetary atmospheres \citep[e.g.,][]{Owen13,Lopez13,Jin14}.
Recently, \cite{Jin17ap} showed that the location of the minimum of the distribution depends on 
the bulk composition of these planets, 
and that the observed location is better explained by their cores being rocky rather than icy.  
A similar deduction was made by \cite{Owen17ap} as well.
Therefore, these studies suggest that most close-in low-mass planets may be 
formed within the snow line 
\footnote{Some of these planets may have low densities. For example, planets d and e of Kepler-444 
might have densities comparable to water $\sim 1\,{\rm g\,cm^{-3}}$ \citep{Mills17}.}. 

The division between rocky planets and planets with extensive gaseous envelopes occurs near 
$1.6R_E$ \citep{Rogers15,Weiss14}. 
From the mass-radius relationship we have adopted (Eq.~\ref{eq_MR}), 
this critical radius corresponds to $4.4M_E$.  
Assuming this is the maximum attainable mass by a rocky planet, 
we can determine the critical disc aspect ratio in our disc model to be $h_g/r = 0.03$, 
where the pebble isolation mass becomes comparable to the critical mass.

\begin{figure*}[ht]
\begin{subfigure}{.5\textwidth}
\includegraphics[width=1.\linewidth]{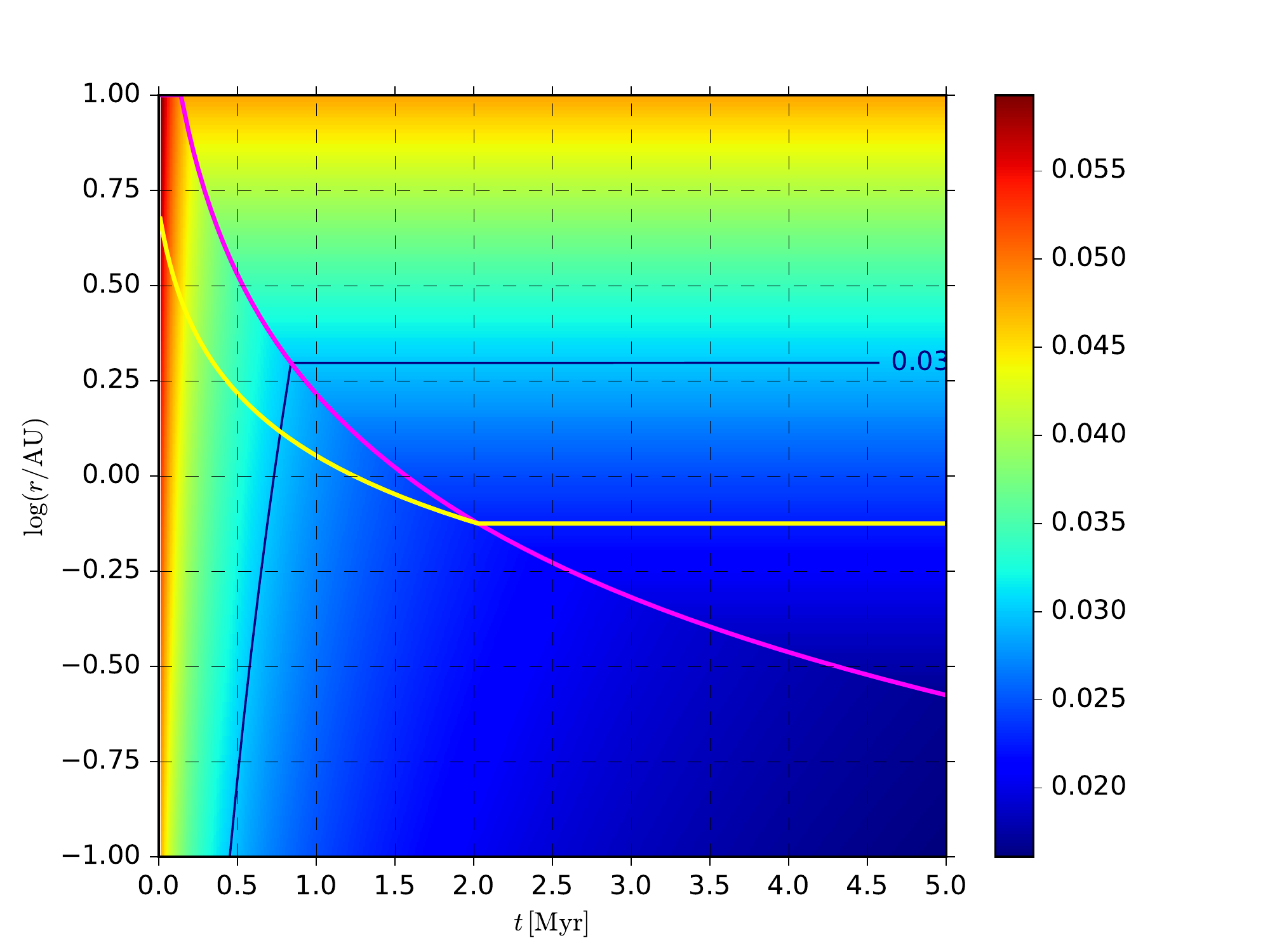}
\end{subfigure}
\begin{subfigure}{.5\textwidth}
\includegraphics[width=1.\linewidth]{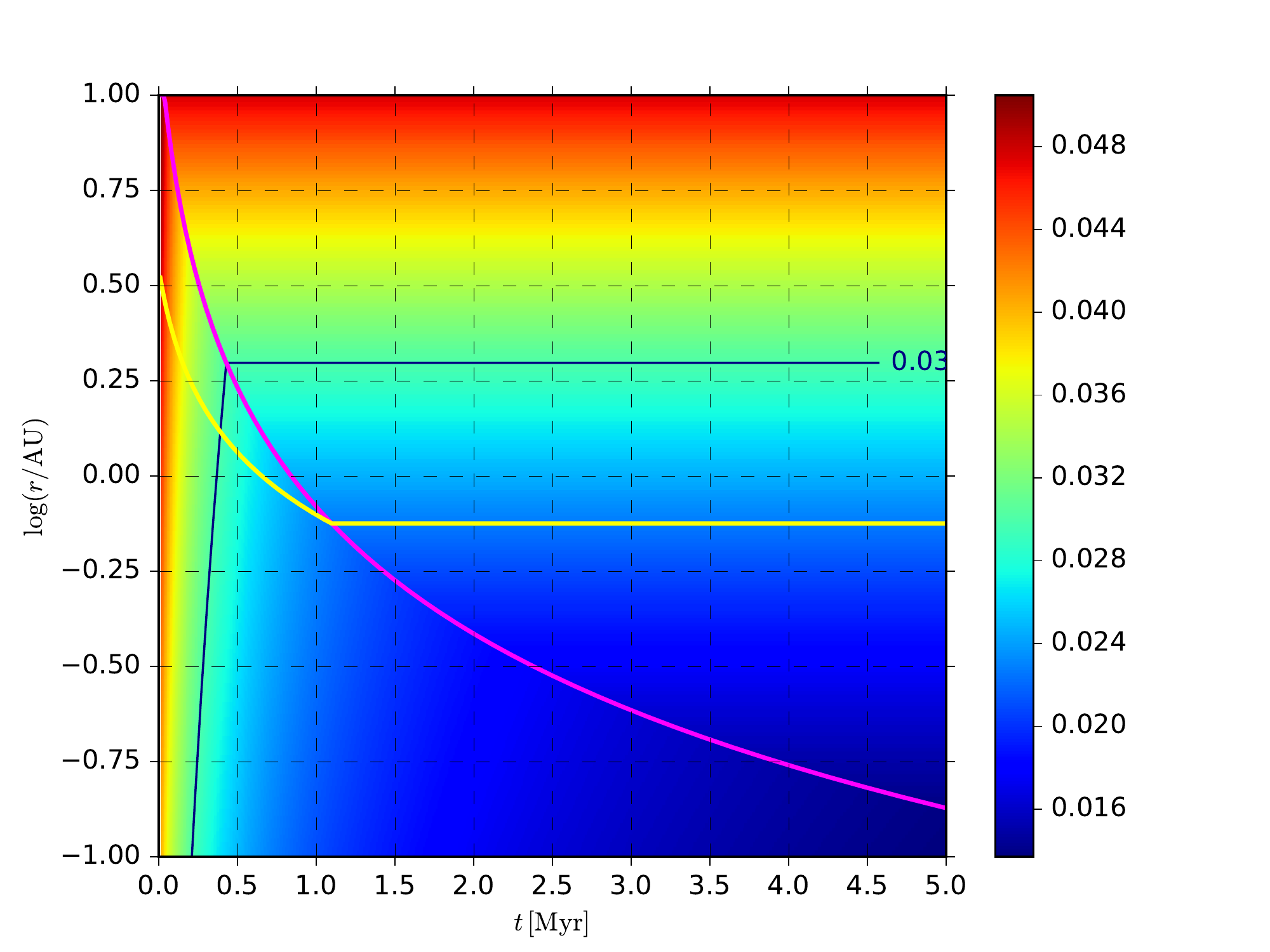}
\end{subfigure}
\caption{The time evolution of a disc aspect ratio of our disc model with $\alpha=10^{-3}$ (left) and $5\times10^{-3}$ (right). 
The black line with 0.03 indicates the critical disc aspect ratio where the pebble isolation mass becomes equal to the critical 
mass for a rocky planet (see text). 
The yellow and magenta lines are the snow line and the viscous-irradiative boundary.  
\label{fig_Ramon}}
\end{figure*}

Figure~\ref{fig_Ramon} plots the time evolution of the disc's aspect ratio 
with $\alpha=10^{-3}$ (left) and $5\times10^{-3}$ (right).
The critical aspect ratio is marked by a solid black line, while yellow and magenta lines 
are the snow line and the boundary between viscous and irradiation regions, respectively.
The figure indicates that at least some of these ``rocky'' planets might have been born 
near or beyond the snow line and thus icy planets, because the critical aspect ratio is often near or beyond 
the snow line and the pebble accretion becomes less efficient inside the snow line.  
This finding disagrees with the recent studies \citep{Fulton17ap,Owen17ap,Jin17ap}, which 
may further imply the need of adopting a more sophisticated disc model.  

Alternatively, some of these low-mass close-in planets could initially have much lower masses since they 
were formed within the snow line and they collided with other planets and grew as the gas disc dissipated.  
Such a scenario is plausible if a chain of planets are formed, for example, 
near the disc's inner edge \citep{Ogihara10}.
%
%
%
\section{Summary}
In this paper, we have investigated global planet formation by considering the pebble accretion 
model by \cite{Ida16a} and compared the outcomes of numerical simulations 
with observed trends of extrasolar planetary systems. 
The N-body code SyMBA has been modified to take account of various effects such as 
pebble accretion, gas accretion, eccentricity and inclination damping 
as well as type I and type II migration (Section~\ref{methods}).
We have performed 135 simulations each, without and with planet migration, 
by varying parameters such as a stellar metallicity ${\rm [Fe/H]}=(-0.5,\,0.0,\,0.5)$, 
a disc's viscosity parameter $\alpha=(10^{-3},\,5\times10^{-3},\,10^{-2})$, and 
an initial disc age $t_{\rm init}=(0.1,\,0.5,\,1.0)\,$Myr. 

Our simulations have confirmed that pebble accretion indeed leads to 
fast formation of giant planets (see Figure~\ref{fig_tgiant}) and 
diverse planetary systems (see Section~\ref{sec_aMIda16}) as shown previously 
\citep[e.g.,][]{Levison15b,Bitsch15,Ida16a,Chambers16}. 
However, the distributions of semimajor axis, eccentricity, 
and planetary mass have not been reproduced (see Section~\ref{sec_aMall}). 
The simulations with migration led to no massive giant planets ($>0.17\,M_J$) and Es and SEs are clustered near the disc edge, 
which is largely due to a too-efficient type I migration (see Section~\ref{sec_typeI}).
Also, our disc edge may have been too wide to trap multiple planets in the resonance 
(see Section~\ref{sec_inneredge}). 
Furthermore, despite that the eccentricity distribution is largely determined by planet-planet interactions 
\citep[e.g.,][]{Ford08,Chatterjee08,Juric08}, 
the eccentricities of our planets are lower than observed ones (see Section~\ref{sec_aMall}).  
This is partly because of a shortage of systems with a large number of giant planets, 
which seems to be caused by the choice of a disc evolution model (see Section~\ref{sec_eM}).

Having these in mind, our simulations have led to the following general trends. 
%
\begin{enumerate}
%
\item {We find that giant planets tend to be formed in low-viscosity or massive discs, 
while Es and SEs are formed more easily (see Section~\ref{sec_met} and Figure~\ref{fig_number}). 
The trend agrees with previous work done by considering the classical planet formation scenario 
\citep[e.g.,][]{Thommes08,Coleman16b}. 
The higher viscosity leads to the faster disc evolution and the lower pebble mass flux, and thus the slower 
planet formation.}

\item {Not only giant planets but Es/SEs show the dependence of formation efficiency on 
stellar metallicities (see Section~\ref{sec_met} and Figures~\ref{fig_number} and \ref{fig_number2}). 
The higher metallicity leads to the faster formation of a larger number of Es/SEs. 
The fraction of systems with no giant planets ($\leq0.1\,M_J$) decreases roughly from 0.8, 0.7, and 0.4 for 
stellar metallicities of ${\rm[Fe/H]}=-0.5$, 0.0, and 0.5, respectively, while a corresponding fraction 
for systems with no Es/SEs ($0.1\,M_E-0.1\,M_J$) decreases roughly from 0.6, 0.4, and 0.2 with metallicities. 
Such a dependence of Es/SEs on a stellar metallicity supports a recent work by \cite{Wang15}, which shows that 
a planet-metallicity correlation exists for Es/SEs as well. 
However, the dependency of metallicity for Es/SEs is subtle, because the final outcomes depend on the timing of 
planet formation as well as the later dynamical evolution of planets.}    

\item {The dynamical evolution of planets alone (without taking account of evaporation effects etc.) naturally 
leads to Es and SEs with various densities (see Section~\ref{sec_composition} and Figure~\ref{fig_massratios}).  
Before the dynamical instability sets in, planets formed in our simulations have similar core-to-envelope mass ratios 
for the same-mass planets.  After collisional events, $10\,M_E$ SEs can have envelope mass fractions 
ranging from less than 1\% to about 50\%.}
%

\item {The amount of masses ejected from the systems or merged with the central stars has 
no correlation with the total masses of survived planets (see Section~\ref{sec_ejemer} and Figure~\ref{fig_memass}).  
Our simulations show that systems with giant planets do not necessarily lose the 
largest amount of masses via ejection or merger with the central star.}

\item {The ejection of giant planets is a rare event, but a significant fraction of ejected planets may be Es/SEs.  
In our simulations, the ratio of the number of ejected low-mass planets with masses $<1.0\,M_E$ and 
that of planets with $1.0\,M_E-0.1\,M_J$ is 0.5.
\cite{Barclay17ap} recently estimated that WFIRST would detect up to 20 Mars-mass planets but few free-floating Earth-mass planets.
However, our results suggest that at least one Earth or more massive free-floating planet may be discovered 
for two Mars-like planets (see Section~\ref{sec_ejemer})}.      
%
%
%
%
\end{enumerate}

In implementing planet migration as in Section~\ref{migmodel}, 
we also show that it is more appropriate to use the semimajor axis evolution time scale $\tau_a=-a/\dot{a}$ 
rather than the ``migration'' time scale $\tau_m=-L/\dot{L}$.  If $\tau_m$ is used in place of $\tau_a$, that could 
lead to an artificial outward migration at high eccentricities ($e>h/r$, see Figure~\ref{fig_tauaem}).

In summary, our simulations with migration highlighted a difficulty of saving type I migrators with a simple disc model 
(see Sections~\ref{sec_aMall} and \ref{sec_mig}). 
Nevertheless, planets formed in the with-migration simulations show orbital period separations that have 
a good agreement with observations (see Section~\ref{sec_pratios} and Figure~\ref{fig_pratios}).
These planets tend to be on nearly circular and coplanar orbits (see Figure~\ref{fig_eccinc}) and 
planetary systems tend to be compact, in a similar manner to Kepler-detected planets. 
The main reason behind this tendency for compact systems is the early occurrence of dynamical instability.  
In simulations with migration, both planet-planet collisions and planet-star mergers occur mostly within the disc's lifetime 
($\lesssim4\,$Myr, see Figure~\ref{fig_cem}).
Therefore, planetary eccentricities and inclinations could be damped after these events by the disc.  
This is contrasted by the outcomes of simulations without migration, where planet ejections, collision, 
and mergers happen most frequently around the disc dissipation times ($\sim2-10\,$Myr).  
Planets formed in these simulations have higher eccentricities and inclinations compared to those in 
the with-migration simulations. 
In the future work, we will consider a more sophisticated disc model and its evolution, and 
improve our gas accretion model, as discussed in Section~\ref{discussion}.
%
\begin{acknowledgements}
We thank Man Hoi Lee for the discussion regarding the implementation of migration in SyMBA, 
and Hal Levison and Katherine Kretke for discussions on implementing pebble accretion in SyMBA. 
We also thank the referee Phil Armitage for his useful comments and 
Tristan Guillot for feedback and editorial assistance. 
SM would like to thank the Earth-Life Science Institute at Tokyo Institute of Technology for 
its hospitality, where part of this work has been done. 
Numerical simulations were in part carried out on the PC cluster at the 
Center for Computational Astrophysics, National Astronomical Observatory of Japan.
RB and SM thank the Daiwa Anglo-Japanese Foundation for its support through a Small Grant.
RB is grateful for continued support from JSPS KAKENHI Grant Number JP16K17662. 
SI is supported by JSPS KAKENHI Grant Number JP15H02065 as well as MEXT KAKENHI Grant Number JPhp170223.
\end{acknowledgements}
\bibliographystyle{aa}
\bibliography{REF}

\end{document}